\documentclass[journal=jacsat,manuscript=article]{achemso}
\usepackage{graphicx} 
\usepackage{siunitx}
\usepackage{amsmath}
\usepackage[version=3]{mhchem}
\usepackage{tikz}
\usepackage{caption}
\usepackage[labelformat=simple]{subcaption}

\usepackage{tabularx}
\usepackage{chapterbib}
\usepackage{textgreek}
\usepackage{enumitem}
\usepackage{multirow}
\usepackage{longtable}
\usepackage{booktabs}

\author{Moritz C. Schmidt}
\affiliation{LMPV - Sustainable Energy Materials Department, AMOLF, Science Park 104, 1098 XG, Amsterdam, The Netherlands}
\author{Agustin O. Alvarez}
\affiliation{LMPV - Sustainable Energy Materials Department, AMOLF, Science Park 104, 1098 XG, Amsterdam, The Netherlands}
\author{Riccardo Pallotta}
\affiliation{Department of Chemistry, University of Pavia, Via T. Taramelli 14, 27100 Pavia, Italy}
\author{Biruk A. Seid}
\affiliation{University Potsdam, Am Neuen Palais 10, 14469 Potsdam}
\author{Jeroen J. de Boer}
\affiliation{LMPV - Sustainable Energy Materials Department, AMOLF, Science Park 104, 1098 XG, Amsterdam, The Netherlands}
\author{Jarla Thiesbrummel}
\affiliation{LMPV - Sustainable Energy Materials Department, AMOLF, Science Park 104, 1098 XG, Amsterdam, The Netherlands}
\author{Felix Lang}
\affiliation{University Potsdam, Am Neuen Palais 10, 14469 Potsdam}
\author{Giulia Grancini}
\affiliation{Department of Chemistry, University of Pavia, Via T. Taramelli 14, 27100 Pavia, Italy}
\author{Bruno Ehrler}
\affiliation{LMPV - Sustainable Energy Materials Department, AMOLF, Science Park 104, 1098 XG, Amsterdam, The Netherlands}
\email{b.ehrler@amolf.nl}

\title{Quantification of mobile ions in perovskite solar cells with thermally activated ion current measurements}

\begin{document}

\begin{abstract}
Mobile ions play a key role in the degradation of perovskite solar cells, making their quantification essential for enhancing device stability. Various electrical measurements have been applied to characterize mobile ions. However, discerning between different ionic migration processes can be difficult. Furthermore, multiple measurements at different temperatures are usually required to probe different ions and their activation energies. Here, we demonstrate a new characterization technique based on measuring the thermally activated ion current (TAIC) of perovskite solar cells. The method reveals density, diffusion coefficient, and activation energy of mobile ions within a single temperature sweep and offers an intuitive way to distinguish mobile ion species. We apply the TAIC technique to quantify mobile ions of \ce{MAPbI3} and triple-cation perovskite solar cells. We find a higher activation energy and a lower diffusion coefficient in the triple-cation devices.  TAIC measurements are a simple yet powerful tool to better understand ion migration in perovskite solar cells.
\end{abstract}

\begin{figure}[h!]
\centering
    \begin{subfigure}[t]{0.3\textwidth}
    \captionsetup{justification=raggedright, singlelinecheck=false, position=top}
        \subcaption{}\label{p5:f1a}
        \vspace{0.3cm}
        \includegraphics[scale=0.75]{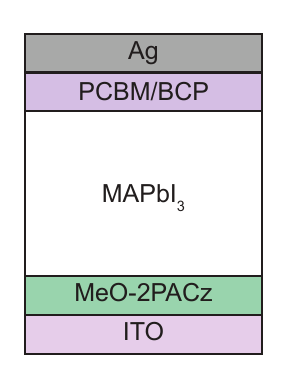}
    \end{subfigure}
    \begin{subfigure}[t]{0.48\textwidth}
        \captionsetup{justification=raggedright, singlelinecheck=false, position=top}
        \subcaption{}\label{p5:f1b}
        \includegraphics[scale=0.9]{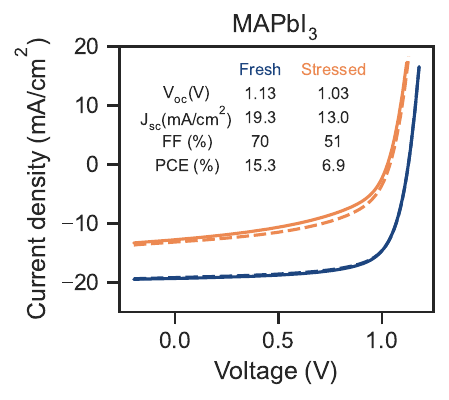}
    \end{subfigure}
    \begin{subfigure}[t]{0.3\textwidth}
        \captionsetup{justification=raggedright, singlelinecheck=false, position=top}
        \subcaption{}\label{p5:f1c}
        \vspace{0.3cm}
        \includegraphics[scale=0.75]{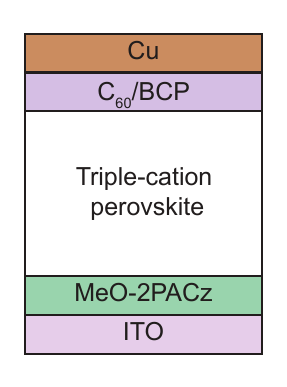}
    \end{subfigure}
    \begin{subfigure}[t]{0.48\textwidth}
        \captionsetup{justification=raggedright, singlelinecheck=false, position=top}
        \subcaption{}\label{p5:f1d}
        \includegraphics[scale=0.9]{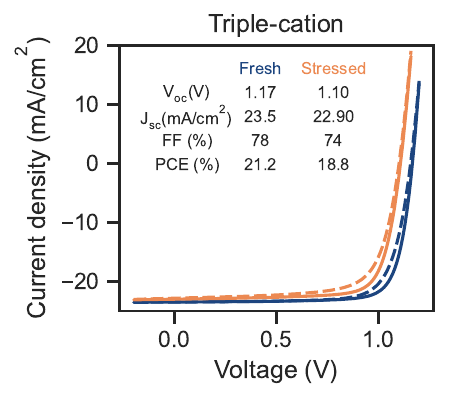}
    \end{subfigure}
    \caption{\subref{p5:f1a} Device stack of the \ce{MAPbI3} perovskite solar cell. \subref{p5:f1b} JV measurement of a fresh \ce{MAPbI3} device and the same device stressed for \SI{32}{h} at $V_{\mathrm{oc}}$. \subref{p5:f1c} Device stack of the triple-cation perovskite solar cell. \subref{p5:f1d} JV measurement of a fresh triple-cation device and the same device stressed for \SI{78}{h} at $V_{\mathrm{oc}}$. The dashed lines are the forward, and the solid lines are the reverse voltage scans. The extracted photovoltaic parameters are the mean values of the forward and reverse measurements.  }
    \label{p5:f1}
\end{figure}

\begin{figure}[hbt]
\centering
        \includegraphics[scale=0.85]{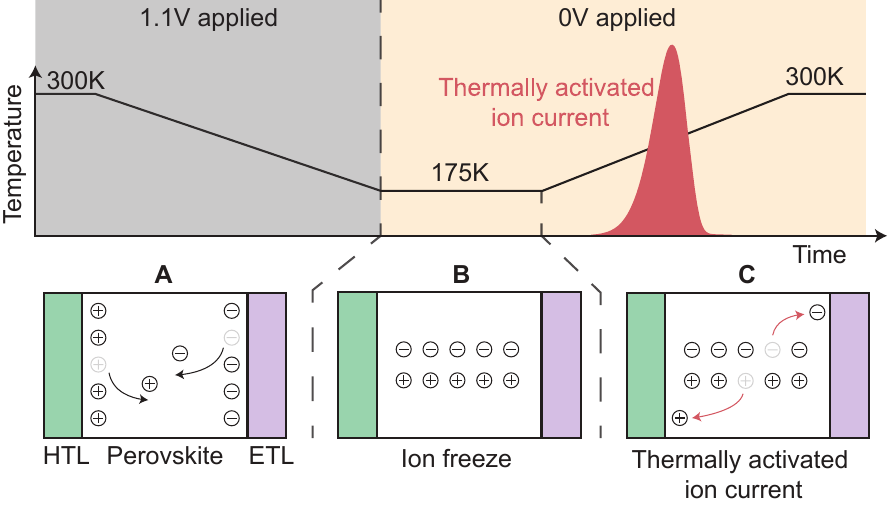}
    \caption{Illustration of the thermally activated ion current measurement. (A) At \SI{300}{K}, a voltage is applied to the device, during which mobile ions migrate into the perovskite bulk. While the voltage is applied, the temperature is decreased, resulting in a decrease in the ionic diffusion coefficient. (B) At \SI{175}{K} the applied voltage is removed. Because of the low temperature, the ions do not drift back to the interface. (C) When the temperature is gradually increased, the diffusion coefficient of the mobile ions increases, resulting in mobile ions drifting back to the perovskite/CTL interfaces, generating the thermally activated ion current (TAIC).}
    \label{p5:f2}
\end{figure}

\begin{figure}[hbt]
\centering
    \begin{subfigure}[t]{0.49\textwidth}
        \captionsetup{justification=raggedright, singlelinecheck=false, position=top}
        \subcaption{}\label{p5:f3a_MAPbI3}
        \includegraphics[scale=0.9]{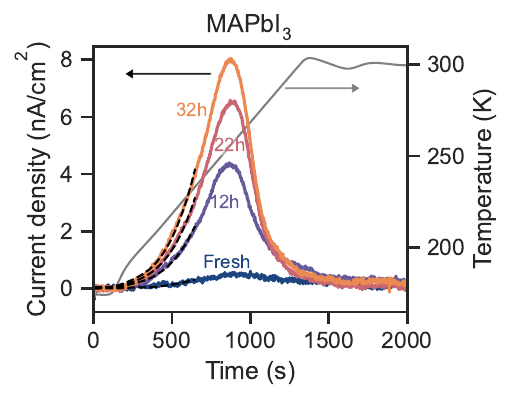}
    \end{subfigure}
    \begin{subfigure}[t]{0.49\textwidth}
        \captionsetup{justification=raggedright, singlelinecheck=false, position=top}
        \subcaption{}\label{p5:f3b_TC}
        \includegraphics[scale=0.9]{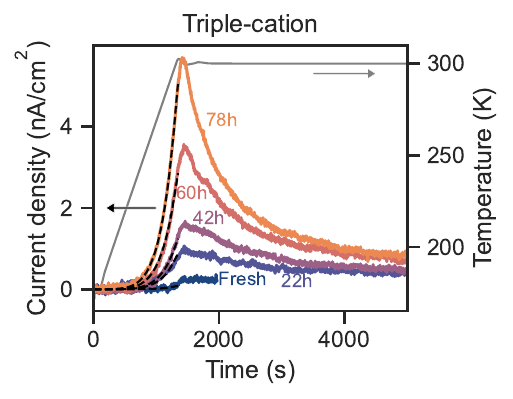}
    \end{subfigure}
    \caption{Thermally activated ion current measurements of \subref{p5:f3a_MAPbI3} a \ce{MAPbI3} and \subref{p5:f3b_TC} a triple-cation perovskite solar cell for different stressing durations. The black dashed lines represent fits. The gray line represents an exemplary temperature sweep. The extracted ion parameters are shown in Table \ref{p5:tab1:IonValues}. }
    \label{p5:f3}
\end{figure}

\begin{figure}[h!]
\centering
    \begin{subfigure}[t]{0.48\textwidth}
        \captionsetup{justification=raggedright, singlelinecheck=false, position=top}
        \subcaption{}\label{p5:fs1a}
        \vspace{-0.3cm}
        \includegraphics[scale=0.85]{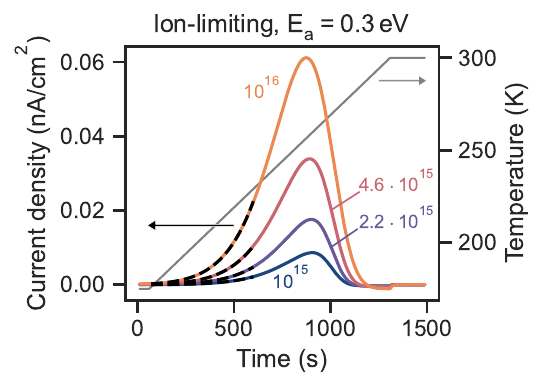}
    \end{subfigure}
    \begin{subfigure}[t]{0.48\textwidth}
        \captionsetup{justification=raggedright, singlelinecheck=false, position=top}
        \subcaption{}\label{p5:fs1b}
        \vspace{-0.3cm}
        \includegraphics[scale=0.85]{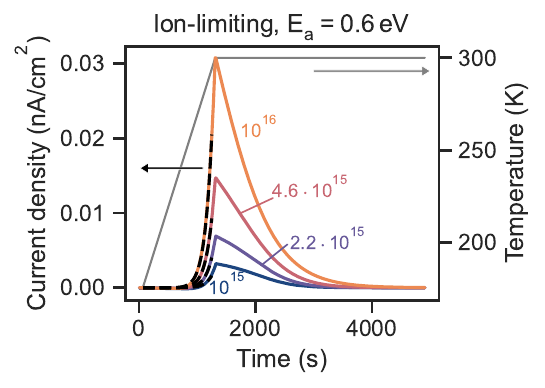}
    \end{subfigure}
    \begin{subfigure}[t]{0.48\textwidth}
        \captionsetup{justification=raggedright, singlelinecheck=false, position=top}
        \subcaption{}\label{p5:fs1c}
        \vspace{-0.3cm}
        \includegraphics[scale=0.85]{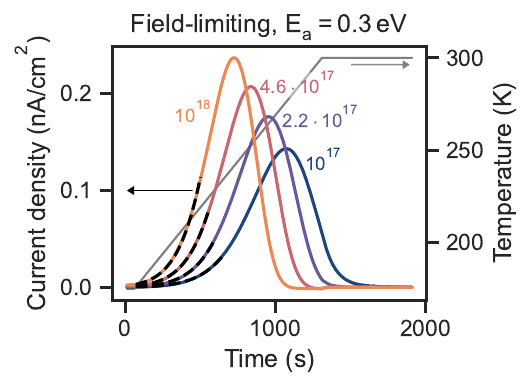}
    \end{subfigure}
    \begin{subfigure}[t]{0.48\textwidth}
        \captionsetup{justification=raggedright, singlelinecheck=false, position=top}
        \subcaption{}\label{p5:fs1d}
        \vspace{-0.3cm}
        \includegraphics[scale=0.85]{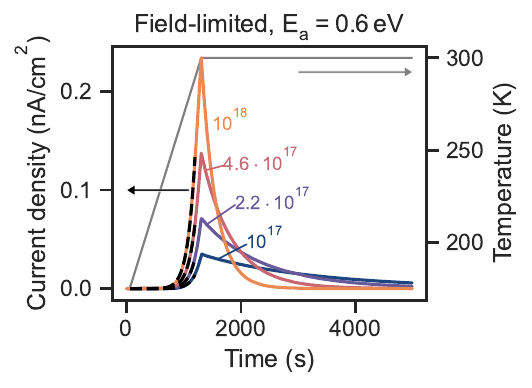}
    \end{subfigure}
    \begin{subfigure}[t]{0.48\textwidth}
        \captionsetup{justification=raggedright, singlelinecheck=false, position=top}
        \subcaption{}\label{p5:fs1e}
        \vspace{-0.3cm}
        \includegraphics[scale=0.85]{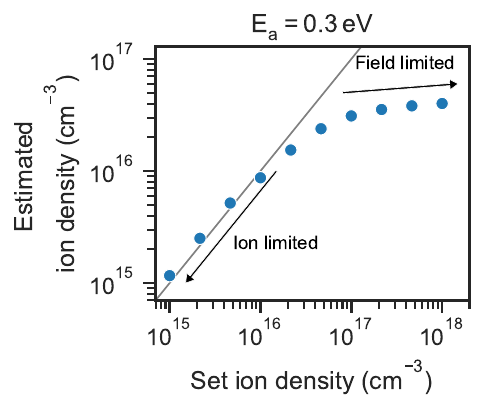}
    \end{subfigure}
    \begin{subfigure}[t]{0.48\textwidth}
        \captionsetup{justification=raggedright, singlelinecheck=false, position=top}
        \subcaption{}\label{p5:fs1f}
        \vspace{-0.3cm}
        \includegraphics[scale=0.85]{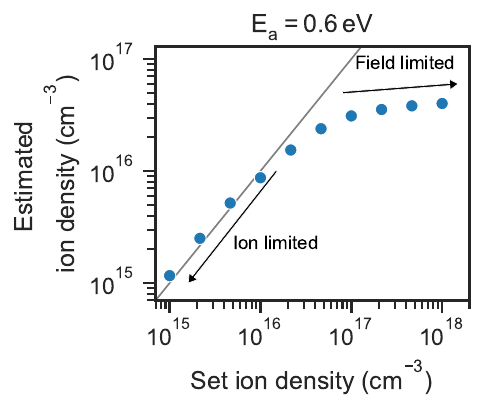}
    \end{subfigure}
    \caption{Drift-diffusion simulations of TAIC measurements with different ion densities for the ion-limiting case with \subref{p5:fs1a} an activation energy of \SI{0.3}{eV} and \subref{p5:fs1b} \SI{0.6}{eV} and the field limiting case with \subref{p5:fs1c} an activation energy of \SI{0.3}{eV} and \subref{p5:fs1d} \SI{0.6}{eV}. All other parameters used for the drift-diffusion simulations are listed in Table \ref{p5:tabS3:DDSimParameters}. The dashed black lines represent fits. \subref{p5:fs1e} Estimated ion density for the simulations with activation energy of \SI{0.3}{eV} and \subref{p5:fs1f} \SI{0.6}{eV}.  }
    \label{p5:fs1}
\end{figure}

In recent years, mobile ions have been assigned to various losses in perovskite solar cells. They have been attributed to losses in short-circuit current density ($J_{\mathrm{sc}}$), open-circuit voltage ($V_{\mathrm{oc}}$), and fill-factor (FF) \cite{thiesbrummel_ion-induced_2024, thiesbrummel_universal_2021, hart_more_2024}. To understand the impact of mobile ions on device characteristics, quantifying key parameters like the ion density, diffusion coefficient, and activation energy is essential. With the aim of extracting these parameters, various electrical measurements have been applied. The density of mobile ions has been estimated with current transient measurements (also known as bias-assisted charge extraction) \cite{thiesbrummel_ion-induced_2024} and low-frequency Mott-Schottky measurements \cite{diethelm_probing_2025, diekmann_determination_2023}. Techniques that are commonly employed to characterize electronic traps in semiconductor materials were transferred to quantify mobile ionic defects in perovskite solar cells, even though their interpretation must be adapted \cite{schmidt_impact_2023}. These techniques include thermal admittance spectroscopy (TAS) \cite{wang_understanding_2018} and deep level transient spectroscopy (DLTS) \cite{lang_deep-level_1974}, which have been employed in efforts to quantify the density, diffusion coefficient, and activation energy of mobile ions \cite{awni_influence_2020, reichert_probing_2020, schmidt_impact_2023, futscher_quantifying_2020, mcgovern_grain_2021}. All of these techniques can be used to quantify the diffusion coefficient and density of ions within some boundary conditions \cite{schmidt_how_2025}. However, multiple measurements at different temperatures are necessary to extract the activation energy of the diffusion coefficient. Additionally, overlapping time constants can make it difficult to discern between different ionic species, especially in transient measurements. \\
Here, we propose an intuitive technique to quantify the density, diffusion coefficient, and activation energy of mobile ions within a single measurement. The method is inspired by thermally stimulated current (TSC) measurements, which have previously been applied to characterize traps in perovskite solar cells \cite{baumann_identification_2015, khan_emergence_2022, gordillo_trap_2017, xu_suppression_2020, leoncini_electronic_2021, moghadamzadeh_spontaneous_2020, ciavatti_radiation_2024}. Similar to current transient measurements, we apply a bias during which mobile ions migrate away from the perovskite/charge transport layer (CTL) interfaces. While applying the bias, we decrease the temperature to \SI{175}{K}, lowering the diffusion coefficient of the mobile ions. At \SI{175}{K}, we then remove the applied bias, resulting in an electric field in the perovskite bulk due to the built-in potential of the device. However, because of the low diffusion coefficient of the mobile ions, they do not immediately drift to the interfaces. Mobile ions only begin to drift back to the perovskite/CTL interface when the temperature is increased at a constant rate, resulting in a thermally activated current. To emphasize that we are probing mobile ionic defects,  we refer to this as the thermally activated ion current (TAIC).  \\
We use the TAIC technique to quantify mobile ions of two different perovskite solar cells, one with a \ce{MAPbI_3} perovskite and one with a triple-cation perovskite. Figure \ref{p5:f1a} and \subref{p5:f1b} show the device stack and JV measurements of the \ce{MAPbI_3} device. In Figure \ref{p5:f1c} and \subref{p5:f1d}, the device stack and JV measurements of the triple-cation device are shown. Importantly, we use optimized device stacks for both the triple-cation \cite{seid_understanding_2024} and the \ce{MAPbI3} \cite{pallotta_reducing_2025} device. This ensures a high shunt resistance, which is necessary to reduce the noise in low-current measurements. Furthermore, we utilize planar CTLs in both devices to avoid complex interface morphologies and to enable modeling of the devices using one-dimensional drift-diffusion simulations. In both devices, we use the self-assembling monolayer MeO-2PACz as the hole transport layer (HTL). For the electron transport layer (ETL), we use PCBM in the case of the \ce{MAPbI_3} perovskite solar cell and \ce{C_{60}} in the case of the triple-cation device. The surface of the triple-cation perovskite is passivated with a dual passivation of EDAI and PEAI. Details of the fabrication process are available in the Experimental Section. \\
It has previously been demonstrated that stressing perovskite solar cells at $V_{\mathrm{oc}}$ can lead to degradation by increased ion densities \cite{thiesbrummel_ion-induced_2024}. We therefore stress the devices at $V_{\mathrm{oc}}$ under a high-intensity white-light LED with 1-sun equivalent carrier excitation. During stressing, we repeatedly carry out electrical measurements, including JV, capacitance frequency, current transient, and TAIC measurements. For the \ce{MAPbI_3} device, for example, we perform measurements of the fresh device and after 12, 22, and 32 hours of stressing at $V_{\mathrm{oc}}$. The JV measurements are measured under illumination, while the capacitance frequency, current transient, and TAIC measurements are carried out in the dark. Figure \ref{p5:f1b} shows the JV measurement of a fresh \ce{MAPbI_3} device and the same device stressed for a total of 32 hours at $V_{\mathrm{oc}}$, resulting in a significant decrease in $V_{\mathrm{oc}}$, $J_{\mathrm{sc}}$, and $\mathrm{FF}$. The triple-cation device is more stable, with a decrease in mainly $V_{\mathrm{oc}}$ and $\mathrm{FF}$ after the maximum stressing time of 78 hours at $V_{\mathrm{oc}}$, as shown in Figure \ref{p5:f1d}.
 \\
We then carry out TAIC measurements of the devices under the different stressing conditions. The principle of the TAIC measurements is illustrated in Figure \ref{p5:f2}. 
At steady state and no applied bias, mobile ions are accumulated at the perovskite/CTL interfaces due to the built-in field of the perovskite. In the first step, we apply a forward bias voltage to the device at \SI{300}{K}. During this applied bias, mobile ions migrate away from the perovskite/CTL interface into the perovskite bulk, as illustrated in panel A in Figure \ref{p5:f2}. 
While still applying the voltage bias, we then decrease the temperature, leading to a decrease in the diffusion coefficient of the mobile ions. Consequently, the ions are 'frozen' when the voltage pulse is removed at \SI{175}{K}, and they do not drift back to the perovskite/CTL interface, as shown in panel B. In the last step, we slowly increase the temperature with the device at short-circuit. As the temperature increases, the temperature-activated diffusion coefficient of the mobile ions increases exponentially. Consequently, mobile ions start drifting back to the perovskite/CTL interfaces. This results in the thermally activated ion current, which is illustrated in panel C in Figure \ref{p5:f2}. 
 \\
The TAIC measurements of the \ce{MAPbI_3} device and the triple-cation device after different stressing durations are shown in Figure \ref{p5:f3a_MAPbI3} and \subref{p5:f3b_TC}, respectively. In both cases, we increase the temperature from \SI{175}{K} to \SI{300}{K} at a rate of \SI{0.1}{K/s} and then stabilize at \SI{300}{K}.  Exemplary temperature sweeps are shown as gray lines in Figure \ref{p5:f3a_MAPbI3} and \subref{p5:f3b_TC}. We expect the \ce{MAPbI3} perovskite to be in the tetragonal phase during the entire temperature sweep \cite{kim_first-principles_2020, whitfield_structures_2016} and see no obvious signs of a phase transition in the triple-cation devices. For both devices, we observe an increase in the current as the temperature increases. At some point, the current peaks and decreases again. For the \ce{MAPbI3} device, this increase and decrease of the current occur during the temperature sweep. In contrast, the current peak in the triple-cation device occurs only when the temperature sweep is stopped at \SI{300}{K}. Notably, in both cases, the integral of the current increases with increasing stressing time, suggesting that stressing the devices at $V_{\mathrm{oc}}$ increases the ion density in both devices. We also measured a second \ce{MAPbI3} and triple-cation device, yielding similar trends but slightly different absolute values as shown in Figure \ref{p5:fs4}. \\
We note that for both devices, some charges are extracted immediately after switching off the voltage at \SI{175}{K} as shown in Figure \ref{p5:fs6a} and \subref{p5:fs6b}. These could be caused by electrical or fast ionic carriers that are still mobile at low temperatures. 
 \\
The current $J_{\mathrm{tot}}$ during the TAIC measurements depends on the density of mobile ions in the bulk $N_{\mathrm{ion,bulk}}$, the temperature activated diffusion coefficient with prefactor  $D_{\mathrm{0,ion}}$ and activation energy $E_{\mathrm{a}}$ \cite{futscher_mixed_2021}, and the electric field in the perovskite bulk $E_{\mathrm{bulk}}$ as: 
\begin{align}
    J_{\mathrm{tot}}(t) =  b\,e^2 N_{\mathrm{ion,bulk}}(t) D_{\mathrm{0,ion}}e^{-\frac{E_{\mathrm{a}}}{k_{\mathrm{B}} T(t)}} \frac{1}{k_{\mathrm{B}} T(t)} \, E_{\mathrm{bulk}}(t) 
    \label{p5:equ2_ionicCurrentComplete}
\end{align}
where $b$ is a correction factor accounting for the displacement current in the perovskite, $e$ is the elementary charge, $k_{\mathrm{B}}$ is the Boltzmann constant, and $T$ is the temperature. A detailed derivation of Equation \ref{p5:equ2_ionicCurrentComplete} is given in Note 1 in the Supporting Information. \\
At low temperatures, we can assume that the density of ions in the bulk is constant $N_{\mathrm{ion,bulk}}(t) = N_{\mathrm{ion}}$, because the bulk is not yet depleted of mobile ions. Furthermore, we can approximate the bulk electric field based on an estimated built-in potential of the devices and a potential drop in the transport layers as described in Note 2 in the Supporting Information. With these simplifications, we can then fit Equation \ref{p5:equ2_ionicCurrentComplete} to the low-temperature part of the TAIC measurements to determine the activation energy and product of ion density and diffusion coefficient $N_{\mathrm{ion}}  D_{\mathrm{0,ion}}$.  \\
To verify this approach, we fit Equation \ref{p5:equ2_ionicCurrentComplete} to drift-diffusion simulations of TAIC measurements, shown in Figure \ref{p5:fs1}. We note that the drift-diffusion solver we use only allows for a temperature-activated mobility instead of the diffusion coefficient. We therefore extract the temperature-independent prefactor of the ionic conductivity $\sigma_{\mathrm{0,ion}} = e N_{\mathrm{ion}} \mu_{\mathrm{0,ion}}$, which we can estimate with good accuracy as shown in Table \ref{p5:tabs1_FittingDD}. \\
For the devices in Figure \ref{p5:f3a_MAPbI3} and \subref{p5:f3b_TC}, we assume a built-in voltage of \SI{1}{V}. With the correction factor described in Note 2 in the Supporting Information, we estimate an electric field of \SI{16}{kV/cm} for the \ce{MAPbI3} device and \SI{12}{kV/cm} for the triple-cation device. The fits of Equation \ref{p5:equ2_ionicCurrentComplete} are illustrated as dashed lines in  Figure \ref{p5:f3a_MAPbI3} and \subref{p5:f3b_TC}. We fit the activation energy as a global parameter and the product of density and diffusion coefficient $N_{\mathrm{ion}} D_{\mathrm{0,ion}}$ as local parameters. For the \ce{MAPbI3} and the triple-cation device, we extract activation energies of \SI{0.28}{eV} and \SI{0.35}{eV}, respectively. The extracted values for $N_{\mathrm{ion}} D_{\mathrm{0,ion}}$ are listed in Table \ref{p5:tabS3:MeasurementDevice1FittingValues} in the Supporting Information. With the activation energy, $N_{\mathrm{ion}} D_{\mathrm{0,ion}}$, and Equations \ref{p5:equ_IonicConductivity}, \ref{p5:equ_IonicMobility}, and \ref{p5:equ_IonicDiffusionCoeff} we can now determine the ionic conductivity at different temperatures, for example \SI{300}{K}, which are listed in Table \ref{p5:tab1:IonValues}. For both devices, we extract an increasing ionic conductivity with increasing stressing duration, most likely caused by an increasing ion density in the stressed devices. Furthermore, due to the higher activation energy, the ionic conductivity at \SI{300}{K} of the triple-cation device is 1-2 orders of magnitude lower compared to the \ce{MAPbI3} device. \\
To extract the density and diffusion coefficient from $N_{\mathrm{ion}} D_{\mathrm{0,ion}}$, we need to determine either the density $N_{\mathrm{ion}}$ or the diffusion coefficient $ D_{\mathrm{0,ion}}$. In principle, the integral of the TAIC current can be used to determine the ion density. This is, however, only possible as long as the electric field within the perovskite bulk in Equation \ref{p5:equ2_ionicCurrentComplete} does not significantly change over time. If the electric field is constant, more and more ions drift from the bulk to the perovskite/CTL interfaces, until the bulk becomes depleted of mobile ions, decreasing the current. Then, the integral of the current can be used to approximate the ion density. Because the depletion of ions in the bulk limits the current, we refer to this case as ion-limited.
\begin{table}[hbt]
\centering
\setlength{\tabcolsep}{8pt} 
\renewcommand{\arraystretch}{1.4} 
{\small
\begin{tabular}{c c c c c c  }
Device & Stressing & $E_{\mathrm{a}}$ (eV) & $\sigma_{\mathrm{ion,300K}}  \mathrm{(S/cm)} $ &  $N_{\mathrm{ion}} \mathrm{(cm^{-3}})$ & $D_{\mathrm{ion,300K}} \mathrm{(cm^2/s)}$ \\ \hline
\multirow{4}{*}{\ce{MAPbI_3}} & Fresh & \multirow{4}{*}{0.28} & $3.9\pm0.3\cdot 10^{-13}$ &  $1.8\pm0.2\cdot 10^{17}$  & $3.5 \pm 0.4 \cdot 10^{-13} $  \\
                & 12h &   & $2.6\pm 0.2 \cdot 10^{-12} $ &  $8.8\pm0.1\cdot 10^{17}$  & $4.9 \pm 0.3 \cdot 10^{-13} $ \\
                & 22h &   & $3.9 \pm 0.2 \cdot 10^{-12} $ & $10.5\pm0.1\cdot 10^{17}$  & $6.0\pm 0.3 \cdot 10^{-13} $  \\
                & 32h &   & $5.1\pm 0.3 \cdot 10^{-12} $ & $13.7\pm0.1\cdot 10^{17}$  & $6.0 \pm 0.3 \cdot 10^{-13} $  \\ \hline
\multirow{5}{*}{Triple-cation} & Fresh & \multirow{5}{*}{0.35} & $1.0\pm0.1 \cdot 10^{-14} $ & $1.9\pm1.2\cdot 10^{17}$ &  $8.7\pm 5.4 \cdot 10^{-15} $  \\
                & 22h &   & $9.7 \pm 0.3 \cdot 10^{-14} $ & $15.9\pm1.2\cdot 10^{17}$ &  $9.8 \pm 0.7 \cdot 10^{-15} $ \\
                & 42h &   & $14.0 \pm 0.5 \cdot 10^{-14} $ & $20.9\pm0.9\cdot 10^{17}$ &  $10.8 \pm 0.6 \cdot 10^{-15} $\\
                & 60h &   & $31.8 \pm 1.0 \cdot 10^{-14} $ & $32.1\pm0.6\cdot 10^{17}$ & $16.0 \pm 0.6 \cdot 10^{-15} $ \\
                & 78h &   & $56.3 \pm 1.9 \cdot 10^{-14} $ & $43.8\pm0.5\cdot 10^{17}$ &   $20.8 \pm 0.7 \cdot 10^{-15} $ \\ \hline
\end{tabular}
}
\caption{Estimated values of the activation energy $E_{\mathrm{a}}$, ionic conductivity at \SI{300}{K} $\sigma_{\mathrm{ion,300K}}$, ion density $N_{\mathrm{ion}}$, and diffusion coefficient at \SI{300}{K} $D_{\mathrm{ion,300K}}$ for the \ce{MAPbI3} and the triple-cation device at different stressing conditions. The values were extracted from the low-temperature fit and the integral of the TAIC measurements. The error of $N_{\mathrm{ion}}$ is estimated from the minimum detectable ion density based on the noise of the current and the diffusion coefficient at the temperature of the current peaks. The errors of the $\sigma_{\mathrm{ion,300K}}$ correspond to the fitting error. The error of $D_{\mathrm{ion,300K}}$ is propagated based on the errors of $N_{\mathrm{ion}}$ and $\sigma_{\mathrm{ion,300K}}$. }
\label{p5:tab1:IonValues}
\end{table} \\
In contrast, the electric field $E_{\mathrm{bulk}}(t)$ in Equation \ref{p5:equ2_ionicCurrentComplete} can also limit the current. This occurs when ions that accumulate at the interface between perovskite and CTLs screen the built-in potential \cite{thiesbrummel_ion-induced_2024, shah_impact_2024}, decreasing the electric field in the perovskite bulk and therefore the current. In this case, only a fraction of ions drift from the bulk to the interfaces, and the ion density is underestimated when integrating the current \cite{diekmann_determination_2023}. We refer to this case as the field-limited case.\\
To illustrate the ion-limited and the field-limited cases, we carried out drift-diffusion simulations of the TAIC measurements for different ion densities and activation energies, which are shown in Figure \ref{p5:fs1}. In the ion limiting case for an activation energy of \SI{0.3}{eV}, the peak of the TAIC measurements does not shift significantly when the ion density increases, as illustrated in Figure \ref{p5:fs1a}. For a higher activation energy of \SI{0.6}{eV} in Figure \ref{p5:fs1b}, the TAIC currents decay similarly fast for the different ion densities. In contrast to these observations stands the field limiting case. Here, the peak shifts to shorter times for an activation energy of \SI{0.3}{eV} due to the earlier screening of the built-in field, as shown in Figure \ref{p5:fs1c}. For an activation energy of \SI{0.6}{eV}, the earlier screening of the built-in field results in faster decays for higher ion densities, as shown in Figure \ref{p5:fs1d}. Based on these observations, TAIC measurements can be used to determine if the device suffers from ionic field screening when devices with different ion densities are measured (e.g., during aging).  \\
In the ion-limited case, most mobile ions drift from the bulk to the perovskite/CTL interfaces. We can therefore estimate the ion density by integrating the overall current according to: 
\begin{align}
    N_{\mathrm{ion}} = b \frac{d_{\mathrm{Pero}}}{2}\int_0^\infty J_{\mathrm{tot}}(t)dt
    \label{p5:equ3_ionicDensity}
\end{align}
where $b$ is a correction factor accounting for the drop of the potential in the CTLs. Note 2 in the Supporting Information contains details about the correction factor. $d_{\mathrm{Pero}}$ is the perovskite thickness, and $J_{\mathrm{tot}}$ is the measured current. The factor $\frac{1}{2}$ originates from the assumption that the mobile ions are distributed homogeneously across the bulk. Then, the average distance that mobile ions migrate is $\frac{d_{\mathrm{Pero}}}{2}$.  When applying Equation \ref{p5:equ3_ionicDensity} to drift-diffusion simulations, we can accurately determine the ion density in the ion-limited case, as shown in Figure \ref{p5:fs1e} and \subref{p5:fs1f} and Table \ref{p5:tabs1_FittingDD} in the Supporting Information. For increasing ion densities in the field limited case, the estimated ion density in Figure \ref{p5:fs1e} and \subref{p5:fs1f} plateaus, and the extracted ion densities are significantly underestimated. This is consistent with the observation that no electrical measurement can accurately extract ion densities in the field-limited case \cite{schmidt_how_2025}. With the ionic conductivity and the density, we can now also determine the mobility of ions in the ion-limited case with good accuracy, as shown in Table \ref{p5:tabs1_FittingDD}. 
 \\
In the measurements in Figures \ref{p5:f3a_MAPbI3} and \subref{p5:f3b_TC}, we do not observe a significant shift of the current peak or a faster decay for the more stressed devices. We can therefore assume that the TAIC current is ion-limited. Consequently, we estimate the ion density by integrating the current according to Equation \ref{p5:equ3_ionicDensity}. We note that the current in the triple-cation device does not fully decay within \SI{5000}{s}. We therefore extrapolate the data with exponential decays. The estimated ion densities for the different stressing conditions are listed in Table \ref{p5:tab1:IonValues}. We determine that ion densities for both devices increase by around one order of magnitude due to stressing. For the \ce{MAPbI3} device, the ion density increases from \SI{1.8e17}{cm^{-3}} to \SI{1.4e18}{cm^{-3}}. Similarly, stressing the triple-cation device increases the ion density from \SI{1.9e17}{cm^{-3}} to \SI{4.4e18}{cm^{-3}}. We can now also determine the diffusion coefficients at \SI{300}{K}, which are listed in Table \ref{p5:tab1:IonValues}. For both devices, the diffusion coefficient increases slightly by a factor of two due to stressing. Notably, the diffusion coefficient and ionic conductivity of the \ce{MAPbI3} device are higher than those of the triple-cation device, suggesting that ion migration in the triple-cation devices is suppressed. \\
We note that, just accounting for the electrostatic effects of mobile ions, we would expect ionic field screening to limit the extracted current for ion densities of \SI{e17}{cm^{-3}} and higher (similar to the drift-diffusion simulations in Figure \ref{p5:fs1}). However, we extract much higher ion density from the TAIC measurements. Possibly, more ions can accumulate at the interface between perovskite and CTLs before the field is screened, which has previously been suggested \cite{jacobs_two_2018}. It is also possible that ions recombine when drifting back to the interfaces, not impacting the potential anymore \cite{birkhold_direct_2018}, or that lateral ion migration \cite{jacobs_lateral_2022} impacts the current, leading to an overestimation of the ion density. Pinpointing the exact cause for the high extracted ion densities is a crucial next step to better understand ion migration, but it is out of the scope of this work.   \\
Interestingly, the extracted diffusion coefficients of both devices are significantly lower than those associated with halide vacancy-mediated ion migration, which is often assumed to be the dominating ionic species. For \ce{MAPbI_3}, diffusion coefficients of around \SI{4e-11}{cm^2/s} up to \SI{e-9}{cm^2/s} have been previously assigned to iodide vacancy migration \cite{li_unravelling_2018, mccallum_bayesian_2024, schmidt_consistent_2024}. For double cation perovskite solar cells, diffusion coefficients were determined to be in the range of \SI{e-10}{cm^2/s} \cite{diethelm_probing_2025}. Possibly, we probe the migration of a slower ionic migration process in the TAIC measurements. To verify this, we carried out capacitance frequency measurements, which allow us to probe faster ionic processes. The resulting capacitance frequency measurements for different stressing times are shown in Figure \ref{p5:fs2}. We observe a rise of the capacitance at around \SI{100}{Hz} for both the \ce{MAPbI_3} and the triple-cation device in Figure \ref{p5:fs2a} and \subref{p5:fs2b}, respectively. A similar capacitance rise has been observed in other capacitance frequency measurements and can be associated with ionic defects \cite{jacobs_two_2018, awni_influence_2020, schmidt_consistent_2024}. Interestingly, the capacitance at lower frequencies increases when the devices are stressed. This suggests the presence of another ionic migration process \cite{jacobs_two_2018}, which is slower than the ionic process at around \SI{100}{Hz}. This process can not be completely resolved with capacitance frequency measurements. However, carrying out capacitance frequency measurements at \SI{360}{K} shifts the fast defect to higher frequencies and also reveals more of the capacitance increase of the slower ionic defect, as shown in Figure \ref{p5:fs2c} and \subref{p5:fs2d} for the \ce{MAPbI3} and the triple-cation device, respectively. Additionally, we extracted some current after switching off the voltage at \SI{175}{K}, shown in the current transient measurements in Figure \ref{p5:fs6}. Possibly, this current is due to the fast defect observed in the capacitance frequency measurements. From these current transient measurements, we estimate the densities of the fast ion migration process for the \ce{MAPbI3} and the triple-cation device to be around \SI{e16}{cm^{-3}} and \SI{5e16}{cm^{-3}}, see Table \ref{p5:tabS5:Nion_LowTtransients}. These low ion densities are difficult to resolve in TAIC measurements. Altogether, we assign the fast process in the capacitance frequency and low temperature current transients to the migration of halide vacancies. In the TAIC measurements, we then measure the slower ionic migration process, which we probe only partially at low frequencies in the capacitance frequency measurements. The slow ionic process could be due to cation vacancies, which have been associated with lower diffusion coefficients \cite{futscher_quantification_2019, eames_ionic_2015}. However, the activation energy associated with cation migration is expected to be around  0.8-\SI{1.2}{eV}\cite{eames_ionic_2015, haruyama_first-principles_2015, jong_influence_2018}, significantly larger than the values found here. A more likely explanation is, therefore, that the fast and slow migration processes are due to different migration pathways of the same ion. It has, for example, been previously reported that ion migration along grain boundaries is significantly faster than migration through perovskite grains \cite{shao_grain_2016, mcgovern_grain_2021, ghasemi_multiscale_2023}. Consequently, the current in the TAIC measurements and the slow process in the capacitance frequency measurements could originate from halide vacancies migrating through perovskite grains, whereas the fast process is caused by halide vacancy migration along grain boundaries. To gain a more detailed mechanistic understanding of the origin of the mobile ions, techniques such as time-of-flight secondary ion mass spectrometry \cite{wei_ion-migration_2018} and Kelvin probe force microscopy \cite{birkhold_direct_2018, weber_how_2018} could further support the TAIC measurements, but are not the focus of this work.  \\
To verify that the TAIC signal is due to ionic carriers, we also carried out the TAIC measurement with an applied voltage of \SI{0}{V} during the cool-down. Then, as shown in Figure \ref{p5:fs3a}, we do not observe any current peak, because no ions were moved into the perovskite bulk. This also verifies that a possible temperature-dependent change of the depletion layers in the CTLs is not the origin of the current. To exclude that we are probing trap emission, we illuminated a device at low temperatures, during which traps would be filled. The emission of electronic charges from these traps would result in a similar current to that with applied bias. However, we do not measure any significant current, as shown in Figure \ref{p5:fs3b}. We can therefore assign the measured TAIC signal to mobile ions in the perovskite. As additional control measurements, we carried out simple transient current measurements at \SI{300}{K}, which are shown in Figure \ref{p5:fs8a} and \subref{p5:fs8b} for the \ce{MAPbI3} and triple-cation device, respectively. The extracted ion densities from the current transients, listed in Table \ref{p5:tabS8:Nion_300Ktransients}, are slightly lower compared to the densities extracted from the TAIC measurements, but follow the same trend, showing higher ion densities for longer stressing times. We attribute the difference in ion density between current transient and TAIC measurements to the shorter measurement duration used for the transient current measurements and the duration of the applied bias, which is longer for the TAIC measurements, possibly resulting in the activation of more ions. \\
Finally, we carried out TAIC measurements starting and finishing at a higher temperature of \SI{360}{K} with the aim of observing additional ionic processes. Figure \ref{p5:fs7a} and \subref{p5:fs7b} show exemplary measurements of a \ce{MAPbI3} and the triple-cation device, respectively. We note that the devices degrade while keeping them at \SI{360}{K} for extended durations, complicating a controlled stressing profile. In the \ce{MAPbI3} device, we observe an additional peak and a shoulder at around \SI{310}{K} and \SI{340}{K}, indicating additional ionic processes. For the triple-cation device in Figure \ref{p5:fs7b}, we observe a distinct second peak for which we can extract an activation energy of \SI{0.94}{eV}. Similarly high activation energies have been previously associated and computationally predicted with the migration of cations in perovskites \cite{eames_ionic_2015, haruyama_first-principles_2015, jong_influence_2018, yang_fast_2016}. \\
These measurements illustrate that we can distinguish between different defects within a single temperature sweep. In other techniques like transient current, transient capacitance, and capacitance frequency measurements, multiple measurements at different temperatures are necessary to probe different defects. And even then, it can be difficult to distinguish between different defects in transient measurement, as their characteristic time constants can overlap.  The main disadvantage of TAIC measurements is, however, that low ion densities cannot easily be resolved (like the fast defect probed in capacitance frequency measurements). However, once the density of ions is high enough, there is no inherent limitation on which perovskite can be probed as long as the field is not screened and the perovskite is not significantly doped. Highly doped perovskites, such as tin-based ones \cite{pitaro_tin_2022}, should be treated with caution, as their potential distribution can be vastly different from that of the devices studied in this work. \\
To clarify the position of TAIC measurements within electrical characterization methods, we summarize the capabilities and limitations of commonly-used techniques in Table \ref{p5:tab2:ComparisonTechniques}. We note that all the methods can only quantify mobile ion densities if mobile ions do not significantly screen the electric field \cite{schmidt_how_2025}.  
\begin{table}[hbt]
\centering
\setlength{\tabcolsep}{5pt} 
\renewcommand{\arraystretch}{1.5} 
{\footnotesize
\begin{tabular}{p{2.5cm} p{6cm} p{6cm}  }
Technique & \qquad Capabilities & \qquad Limitations  \\ \hline
Thermally activated ion current (TAIC) &
\begin{minipage}[t]{\linewidth}
\begin{itemize}[nosep]
\item Measurement of ion density and diffusion coefficient 
\item Single temperature sweep to determine activation energy 
\item Intuitive measurement to distinguish between different ions
\end{itemize}
\end{minipage}
 & 
 \begin{minipage}[t]{\linewidth}
 \begin{itemize}[nosep]
    \item Difficult to measure low ion densities
    \end{itemize}
    \end{minipage}
 \\ \hline
 Transient current \cite{thiesbrummel_ion-induced_2024, bertoluzzi_mobile_2020, diethelm_probing_2025, alvarez_ion_2024} &
\begin{minipage}[t]{\linewidth}
\begin{itemize}[nosep]
\item Measurement of ion density and diffusion coefficient 
\item Slow and fast ions can be measured 
\item Sensitive to low ion densities
\end{itemize}
\end{minipage}
 & 
 \begin{minipage}[t]{\linewidth}
 \begin{itemize}[nosep]
    \item Difficult to distinguish between different ions
    \item Multiple measurements at different temperatures are necessary to determine the activation energy 
    \end{itemize}
    \end{minipage}
 \\ \hline
 Transient capacitance \cite{futscher_quantification_2019, reichert_probing_2020, mcgovern_reduced_2021, schmidt_consistent_2024} &
\begin{minipage}[t]{\linewidth}
\begin{itemize}[nosep]
\item Measurement of ion density and diffusion coefficient 
\item Slow and fast ions can be measured 
\item Sensitive to low ion densities
\end{itemize}
\end{minipage}
 & 
 \begin{minipage}[t]{\linewidth}
 \begin{itemize}[nosep]
    \item A more complex model is necessary to evaluate the data
    \item Difficult to distinguish between different ions
    \item Multiple measurements at different temperatures are necessary to determine the activation energy 
    \end{itemize}
    \end{minipage}
 \\ \hline
Capacitance frequency \cite{awni_influence_2020, jacobs_two_2018, schmidt_consistent_2024} &
\begin{minipage}[t]{\linewidth}
\begin{itemize}[nosep]
\item Measurement of ion density and diffusion coefficient 
\item Sensitive to low ion densities
\end{itemize}
\end{minipage}
 & 
 \begin{minipage}[t]{\linewidth}
 \begin{itemize}[nosep]
    \item A more complex model is necessary to evaluate the data
    \item Measurement of slow ions takes a long time
    \item Multiple measurements at different temperatures are necessary to determine the activation energy 
    \end{itemize}
    \end{minipage}
 \\ \hline
Low-frequency Mott-Schottky \cite{diekmann_determination_2023, diethelm_probing_2025}  &
\begin{minipage}[t]{\linewidth}
\begin{itemize}[nosep]
\item Measurement of ion density 
\item Sensitive to low ion densities
\end{itemize}
\end{minipage}
 & 
 \begin{minipage}[t]{\linewidth}
 \begin{itemize}[nosep]
    \item Multiple ions can not be measured/distinguished
    \item Does not contain information about the diffusion coefficient
    \end{itemize}
    \end{minipage}
 \\ 
\end{tabular}
}
\caption{Comparison of capabilities and limitations of common electrical measurement techniques used to quantify mobile ions in perovskite solar cells.}
\label{p5:tab2:ComparisonTechniques}
\end{table} \\
In summary, we have introduced a new measurement technique, thermally activated ion current (TAIC), to characterize mobile ions in perovskite solar cells. TAIC is based on measuring the current due to thermally activated ions. With a simple expression for the TAIC current, we extracted the ionic conductivity by fitting the low-temperature tail of the TAIC measurements. Furthermore, the ion density can be determined by integrating the current in the TAIC measurements, if electric field screening does not limit the overall current. Conveniently, the peak shift in the TAIC data indicates if perovskite solar cells suffer from electric field screening. We applied TAIC measurements to quantify the mobile ion density, diffusion coefficient, and activation energy in a \ce{MAPbI3} and a triple-cation perovskite solar cell at different stressing conditions. For the \ce{MAPbI3} device we determined an activation energy of \SI{0.28}{eV}, mobile ion densities of \SI{1.8e17}{cm^{-3}} to \SI{1.4e18}{cm^{-3}} depending on the stressing condition and a diffusion coefficient of around \SI{5e-13}{cm^2/s} at \SI{300}{K}. For the triple-cation device we determined an activation energy of \SI{0.35}{eV}, mobile ion densities of \SI{1.9e17}{cm^{-3}} to \SI{4.4e18}{cm^{-3}}, and a diffusion coefficient of around \SI{e-14}{cm^2/s} at \SI{300}{K}, lower than that of the \ce{MAPbI3} device. We attribute the migration process to halide vacancy migration within perovskite grains. We also observed a faster ionic process in capacitance frequency measurements, which we assign to halide vacancy migration along grain boundaries. Lastly, we showed that it is possible to distinguish between different ionic processes by increasing the temperature range of the TAIC measurements and found a third ion migration process with a high activation energy of \SI{0.94}{eV} in the triple-cation devices, which we assign to cation migration. In total, TAIC measurements are a promising technique because they are easy to perform, their interpretation is straightforward, and they offer an intuitive visualization of ion migration in perovskite solar cells.
\section{Experimental}
\subsection*{Fabrication of the \ce{MAPbI3} devices}

The \ce{MAPbI3} devices were prepared following the procedure described in Pallotta et al. \cite{pallotta_reducing_2025}. \\

\noindent \textbf{Materials:} Chlorobenzene (CB, extra dry, 99.8\%), dimethyl sulfoxide (DMSO, $\geq$99.9\% extra dry), N,N-dimethylformamide (DMF, extra dry, 99.8\%), and chloroform (CF, extra dry 99.8\%) were purchased from Acros Organics. 2-propanol (IPA, $\geq$ 99.8\%), lead iodide (\ce{PbI2}, $>$98.0\%), and MeO-2PACz were purchased from TCI. Methylammonium iodide (MAI, $>$99.99\%) was purchased from GreatCell Solar Materials. Phenyl-C61-butyric acid methyl ester (PCBM, $>$99.99\%) was purchased from Lumatec. Bathocuproine (BCP) was purchased from Sigma Aldrich. All solutions were prepared in an Ar-filled glovebox, while the deposition of each layer of the solar cell was performed in an \ce{N2}-filled glovebox. 

\noindent \textbf{Device Fabrication:}  For the fabrication of the \ce{MAPbI3} devices, indium tin oxide (ITO)-coated glass substrates (purchased from Yingkou Shangneng Photoelectric material Co.,Ltd.) were consecutively cleaned in acetone and IPA by ultrasonicating for \SI{15}{min} in each solvent. Substrates were dried with \ce{N2} airflow and \ce{O2} plasma treated for \SI{10}{min}. MeO-2PACz was dissolved in ethanol in a concentration of \SI{0.33}{mg/ml} and \SI{50}{\micro l} were spin-coated onto ITO/glass substrates at \SI{3000}{rpm} for \SI{30}{s} and annealed at \SI{100}{\degree C} for \SI{10}{min}. The perovskite precursor solution was prepared by dissolving \SI{0.553}{g} of \ce{PbI2} and \SI{0.191}{g} of \ce{MAI} powders in \SI{1}{ml} DMF/DMSO 4/1 v/v solvent. \SI{25}{\micro l} of the final solution were deposited on the MeO-2PACz coated substrates and spin-coated with a three-step procedure: the first step proceeded at \SI{1000}{rpm} (\SI{500}{rpm/s}) for \SI{6}{s}, the second step proceeded at \SI{5000}{rpm} for \SI{27}{s} (\SI{2500}{rpm/s}), while the last step was a speed deceleration of \SI{1250}{rpm/s} to \SI{0}{rpm/s}. \SI{150}{\micro l} of chlorobenzene were dropped onto the spinning substrate for an antisolvent procedure 6 seconds after the beginning of the second step. Subsequently, substrates were annealed at \SI{100}{\degree C} for \SI{15}{min}. To fabricate the ETL, PCBM was dissolved in chloroform to produce a \SI{15}{mg/ml} solution. \SI{20}{\micro l} of the solution were spin-coated at \SI{2000}{rpm} for \SI{20}{s} onto the perovskite layer. To prevent the diffusion of the metal contact into the perovskite, \SI{50}{\micro l} of \SI{1}{mg/ml} solution (in isopropanol) of bathocuproine was deposited on PCBM. For all the deposition, a vacuum-based chuck was used. Finally, \SI{80}{nm} of Ag was thermally evaporated on the device with a shadow mask of \SI{0.0825}{cm^2} area. The evaporation speed was adjusted to \SI{0.01}{nm/s} for the first \SI{5}{nm}, \SI{0.02}{nm/s} from 5 to \SI{15}{nm}, and \SI{0.06}{nm/s} for the rest of the procedure. 

\subsection*{Fabrication of the triple-cation devices}

\textbf{Solution preparation:} The perovskite solution was prepared by adopting the procedure reported by Seid et al. \cite{seid_understanding_2024}.  \ce{PbI2} (\SI{909.00}{mg}), \ce{FAI} (\SI{276.06}{mg}), \ce{MABr} (\SI{3.68}{mg}), \ce{CsI} (\SI{22.47}{mg}), and \ce{MACl} (\SI{18.11}{mg}) were mixed in a DMF/DMSO solvent mixture (5/1 v/v) and stirred for 4 hours at \SI{60}{\degree C} to form a \SI{1.73}{M}  \ce{Cs_{0.05}(MA_{0.05}FA_{0.95})_{0.95}Pb(I_{0.95}Br_{0.05})_{3}} perovskite solution. The passivation layers were prepared using high-purity materials from Sigma-Aldrich: PEAI (98\%) and \ce{EDAI2} ($>$98\%). \SI{3.5}{mg} of PEAI was dissolved in \SI{1}{ml} of isopropanol (IPA), sonicated for 30 minutes. The \ce{EDAI2} solution was prepared by dissolving 2 mg of \ce{EDAI2} in a \SI{2}{ml} 1:1 (v/v) mixture of IPA and toluene  \\

\noindent \textbf{Device fabrication:}
Planar inverted perovskite solar cells were fabricated using the following layer structure: glass/ITO/MeO-2PACz/\ce{Cs_{0.05}(MA_{0.05}FA_{0.95})_{0.95}Pb(I_{0.95}Br_{0.05})_{3}}/\ce{C_{60}}/BCP/Cu. The fabrication started with ITO-coated glass substrates, which were cleaned sequentially in an ultrasonic bath using acetone, Hellmanex (\SI{3}{\%} in deionized water), deionized water, ethanol, acetone, and isopropanol, with each solvent being used for 15 minutes. The cleaned substrates were then exposed to ultraviolet ozone for 30 minutes before being placed in a nitrogen-filled glovebox. \\
Next, a MeO-2PACz layer was spin-coated from a 1 \ce{mmol\, ml^{-1}} ethanol solution at 3000 rpm for 30 seconds, followed by annealing at \SI{100}{\degree C} for 10 minutes. Once the substrates had cooled to room temperature, a triple-cation perovskite solution was spin-coated at 4000 rpm for 40 seconds with a 5-second acceleration time. 7 seconds before the end of the spin-coating process, \SI{300}{\micro l} of chlorobenzene was added as an antisolvent, and the perovskite film was annealed at \SI{100}{\degree C} for 1 hour. For the bi-layered passivation, the \ce{EDAI2} solution was spin-coated onto the perovskite at 5000 rpm for 40 seconds, and annealed at 100°C for 10 minutes. Then, the PEAI solution was spin-coated onto the cooled sample at 5000 rpm for 40 seconds. Afterward, the samples were transferred to an evaporation chamber where \SI{30}{nm} of \ce{C_{60}} was deposited at \SI{0.3}{\angstrom / s}, followed by \SI{8}{nm} of BCP and \SI{100}{nm} of copper, which were evaporated at \SI{0.3}{\angstrom /s} and \SI{0.6}{\angstrom / s}, respectively, under a high vacuum of \SI{e{-7}}{mbar}.

\subsection*{Electrical characterization}
All electrical measurements were carried out in a Janis VPF-100 liquid nitrogen cryostat. During the measurements, the pressure inside the cryostat was around \SI{5e-6}{mbar}. \\
\noindent \textbf{JV measurements:} JV measurements were carried out with an Agilent B2902A source-measure unit and a SOLIS-3C high-power white-light LED from Thorlabs. The intensity of the LED was set so that the short-circuit current density of the devices matched with a JV measurement at AM-1.5G illumination (carried out with a Pico solar simulator by G2V inside a \ce{N2} filled glovebox). \\
\noindent \textbf{TAIC measurements:} TAIC measurements were carried out using an Agilent B2902A source-measure unit. At \SI{300}{K} or \SI{360}{K}, a voltage of \SI{1.1}{V} was applied to the devices. After \SI{60}{s}, and while still applying the bias, the devices were cooled down and stabilized at \SI{175}{K}. When switching off the voltage, the current transient was measured. Then, the temperature was increased to \SI{300}{K} or \SI{360}{K} and stabilized there, while the current was constantly recorded. \\
\noindent \textbf{Transient current measurements:} Transient current measurements at \SI{300}{K} were carried out using an Agilent B2902A source-measure unit. At \SI{300}{K}, a voltage of \SI{1.1}{V} was applied to the devices. After \SI{60}{s}, the voltage was removed, and the current at \SI{0}{V} was recorded. \\
\noindent \textbf{Capacitance frequency measurements:} Capacitance frequency measurements were carried out with the MFIA by Zurich Instruments with an AC amplitude of \SI{20}{mV} in a frequency range of \SI{0.1}{Hz}-500 kHz.

\subsection*{Thickness measurements}
The perovskite film thickness was determined by scratching the films with tweezers and measuring the depth of the scratch with a KLA Tencor P-7 Stylus Profiler.

\subsection*{Drift-diffusion simulations}
Drift-diffusion simulations were carried out with Setfos by Fluxim, and the parameter set listed in Table \ref{p5:tabS3:DDSimParameters}.

\begin{acknowledgement}
The work of M.C.S., A.O.A., J.J.dB., and B.E. received funding from the European Research Council (ERC) under the European Union’s Horizon 2020 research and innovation programme under Grant Agreement No. 947221. The work is part of the Dutch Research Council and was performed at the AMOLF research institute. R.P. and G.G. are thankful to The Ministero dell’Università e della Ricerca (MUR), University of Pavia through the program “Dipartimenti di Eccellenza 2023–2027” for fundings and FARE (Framework per l’Attrazione e Il Rafforzamento delle Eccellenze per la ricerca in Italia) Project EXPRESS (exploring photoferroelectricity in halide perovskites for optoelectronics) “Development and characterization of halide perovskites for photoferroelectrics applications”. B.A.S and F.L. received funding from the Volkswagen Foundation via the Freigeist Program. M.C.S. conceived the work, carried out the simulations and measurements, performed the analysis, interpreted the results, and wrote the manuscript. A.O.A. helped with discussions and commented on the manuscript. R.P. fabricated the \ce{MAPbI3} devices and commented on the manuscript. B.A.S fabricated the triple-cation devices and commented on the manuscript. J.J.dB. carried out thickness measurements and commented on the manuscript. J.T. helped with discussions. F.L. supervised the work of B.A.S and commented on the manuscript. G.G. supervised the work of R.P. and commented on the manuscript. B.E. conceived and supervised the work, interpreted the results, and commented on the manuscript.
\end{acknowledgement}

\begin{suppinfo}
The parameter set for the drift-diffusion simulations, a note about the correction factor accounting for potential drops in charge transport layers, additional measurements, and extracted ionic parameters are provided in the Supporting Information. 
\end{suppinfo}

\newpage
\bibliographystyle{achemso}
\bibliography{article}
\newpage

\setcounter{figure}{0}
\renewcommand{\figurename}{Figure}
\renewcommand{\thefigure}{S\arabic{figure}}

\setcounter{table}{0}
\renewcommand{\tablename}{Table}
\renewcommand{\thetable}{S\arabic{table}}

\setcounter{equation}{0}
\renewcommand{\theequation}{S\arabic{equation}}

\setcounter{page}{1}

\begin{longtable}{ c | c  }
\caption{Parameters used for the drift-diffusion simulations. } 
\label{p5:tabS3:DDSimParameters}
\endfirsthead
\hline
\textbf{Parameter} & \textbf{Value} 
\endhead
\hline
\hline
\textbf{Parameter} & \textbf{Value}   \\ 
\hline 
Band gap perovskite  $E_{\mathrm{g,Pero}}$ (eV) & \SI{1.6}{}   \\ \hline
Electron affinity perovskite $E_{\mathrm{aff,Pero}}$ (eV) &  \SI{3.9}{}    \\ \hline
Dielectric constant perovskite $\epsilon_{\mathrm{r,Pero}}$ & \SI{50}{}   \ \\ \hline
Thickness perovskite $d_{\mathrm{Pero}}$ (nm) & \SI{500}{}  \ \\ \hline
Effective density of states conduction band perovskite $N_{\mathrm{0,CB,Pero}} \mathrm{\,(cm^{-3})}$ & \SI{2.1e18}{}   \\ \hline
Effective density of states valence band perovskite $N_{\mathrm{0,VB,Pero}} \mathrm{\,(cm^{-3})}$ & \SI{2.1e18}{}  \\ \hline
Mobility electrons in perovskite $\mu_{\mathrm{n,Pero}} \mathrm{\,(cm^{2}/Vs)}$ & 1   \\ \hline
Mobility holes in perovskite $\mu_{\mathrm{p,Pero}} \mathrm{\,(cm^{2}/Vs)}$ & 1  \\ \hline
Mobile positive ion density in perovskite  $N_{\mathrm{ion}} \mathrm{\,(cm^{-3})}$ & variable  \\ \hline
Immobile negative ion density  $N_{\mathrm{nion}} \mathrm{\,(cm^{-3})}$ & variable  \\ \hline
Prefactor of mobility of ions $\mu_{\mathrm{0,ion}} \mathrm{\,(cm^{2}/Vs)}$ & variable   \\ \hline
Activation energy of mobility of ions  $E_{\mathrm{a}} \mathrm{\,(eV)}$ & variable   \\  \hline
Band gap HTL $E_{\mathrm{g,HTL}}$ (eV) & \SI{1.9}{}   \\ \hline
Electron affinity HTL $E_{\mathrm{aff,HTL}}$ (eV) &  \SI{3.4}{}    \\ \hline
Dielectric constant HTL $\epsilon_{\mathrm{r,HTL}}$ & \SI{4.0}{}   \ \\ \hline
Thickness HTL $d_{\mathrm{HTL}}$ (nm) & \SI{3}{}  \ \\ \hline
Effective density of states conduction band HTL $N_{\mathrm{0,CB,HTL}} \mathrm{\,(cm^{-3})}$ & \SI{2.1e18}{}   \\ \hline
Effective density of states valence band HTL $N_{\mathrm{0,VB,HTL}} \mathrm{\,(cm^{-3})}$ & \SI{2.1e18}{}  \\ \hline
Mobility holes in HTL $\mu_{\mathrm{p,HTL}} \mathrm{\,(cm^{2}/Vs)}$ & \SI{e-4}{}   \\ \hline
Acceptor doping density in HTL  $N_{\mathrm{A,HTL}} \mathrm{\,(cm^{-3})}$ & \SI{0}{}  \\ \hline
Band gap ETL $E_{\mathrm{g,ETL}}$ (eV) & \SI{2.0}{}   \\ \hline
Electron affinity ETL $E_{\mathrm{aff,ETL}}$ (eV) &  \SI{4.0}{}    \\ \hline
Dielectric constant ETL $\epsilon_{\mathrm{r,ETL}}$ & \SI{5.0}{}   \ \\ \hline
Thickness ETL $d_{\mathrm{ETL}}$ (nm) & \SI{50}{}  \ \\ \hline
Effective density of states conduction band ETL $N_{\mathrm{0,CB,ETL}} \mathrm{\,(cm^{-3})}$ & \SI{2.1e18}{}   \\ \hline
Effective density of states valence band ETL $N_{\mathrm{0,VB,ETL}} \mathrm{\,(cm^{-3})}$ & \SI{2.1e18}{}  \\ \hline
Mobility electrons in ETL $\mu_{\mathrm{n,ETL}} \mathrm{\,(cm^{2}/Vs)}$ & \SI{e-4}{}   \\ \hline
Donor doping density in ETL  $N_{\mathrm{D,ETL}} \mathrm{\,(cm^{-3})}$ & \SI{1e17}{} \\ \hline
Work function anode  $W_{\mathrm{f,anode}} \mathrm{\,(eV)}$ & 5.1 \\ \hline
Work function cathode  $W_{\mathrm{f,cathode}} \mathrm{\,(eV)}$ & 4.1 \\  \hline
Applied voltage before TAIC simulation $V_{\mathrm{app}}$ (V) & \SI{1.1}{}  \\ \hline \hline
\end{longtable}

\section*{Supplementary Note 1: TAIC current}
Generally, the ionic current $J_{\mathrm{ion}}$ can be expressed in terms of the ionic conductivity $\sigma_{\mathrm{ion}}$ and the electric field in the perovskite bulk $E_{\mathrm{bulk}}$ as:
\begin{align}
    J_{\mathrm{ion}} (t) =\, \sigma_{\mathrm{ion}}(t) \, E_{\mathrm{bulk}}(t) 
    \label{p5:equ1_ionicCurrent}
\end{align}
The ionic conductivity is dependent on the ion density $N_{\mathrm{ion,bulk}}$ and mobility of the mobile ions $\mu_{\mathrm{ion}}$: 
\begin{align}
    \sigma_{\mathrm{ion}} (t) = e \, N_{\mathrm{ion,bulk}}(t) \, \mu_{\mathrm{ion}}(t)
    \label{p5:equ_IonicConductivity}
\end{align}
where $e$ is the elementary charge. According to the Nernst-Einstein relation, the mobility depends on the ionic diffusion coefficient $D_{\mathrm{ion}}$ as\cite{futscher_mixed_2021}: 
\begin{align}
    \mu_{\mathrm{ion}}(t) = \frac{D_{\mathrm{ion}}(t) \,e}{k_{\mathrm{B}} T(t)}
    \label{p5:equ_IonicMobility}
\end{align}
where $k_{\mathrm{B}}$ is the Boltzmann constant and $T(t)$ is the temperature at time $t$. The diffusion coefficient of mobile ions in perovskites is a temperature-activated process, following \cite{futscher_mixed_2021}: 
\begin{align}
    D_{\mathrm{ion}}(t) = D_{\mathrm{0,ion}} e^{-\frac{E_{\mathrm{a}}}{k_{\mathrm{B}} T(t)}}
    \label{p5:equ_IonicDiffusionCoeff}
\end{align}
where $D_{\mathrm{0,ion}}$ is a temperature independent prefactor, and $E_{\mathrm{a}}$ is the activation energy associated with the diffusion coefficient. With these relationships, we can define the ionic current in Equation  \ref{p5:equ1_ionicCurrent} in terms of the mobile ion density and diffusion coefficient:
\begin{align}
    J_{\mathrm{ion}} (t) = e^2 N_{\mathrm{ion,bulk}}(t) D_{\mathrm{0,ion}}e^{-\frac{E_{\mathrm{a}}}{k_{\mathrm{B}} T(t)}} \frac{1}{k_{\mathrm{B}} T(t)} \, E_{\mathrm{bulk}}(t) 
\end{align}
Finally, the extracted current is the sum of the ionic current and the displacement current. If potential drops in the CTLs, the displacement current results in a lower total current $J_{\mathrm{tot}}$ compared to the ionic current $J_{\mathrm{ion}}$. We can account for the impact of the displacement current with the correction factor $b$:
\begin{align}
    J_{\mathrm{tot}}(t) = b\,J_{\mathrm{ion}} (t) = b\,e^2 N_{\mathrm{ion,bulk}}(t) D_{\mathrm{0,ion}}e^{-\frac{E_{\mathrm{a}}}{k_{\mathrm{B}} T(t)}} \frac{1}{k_{\mathrm{B}} T(t)} \, E_{\mathrm{bulk}}(t) 
\end{align}
Details about the correction factor are discussed in Supplementary Note 2.

\section*{Supplementary Note 2: Correction factor }
\label{p5:secSI_correctionFactor}
\begin{figure}[hbt]
\centering
    \begin{subfigure}[t]{0.48\textwidth}
        \captionsetup{justification=raggedright, singlelinecheck=false, position=top}
        \subcaption{}\label{p5:fs5a}
        \includegraphics[scale=0.9]{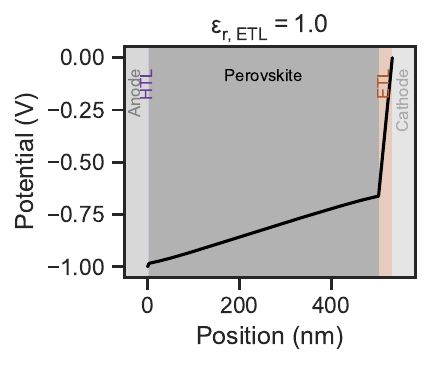}
    \end{subfigure}
    \begin{subfigure}[t]{0.48\textwidth}
        \captionsetup{justification=raggedright, singlelinecheck=false, position=top}
        \subcaption{}\label{p5:fs5b}
        \includegraphics[scale=0.9]{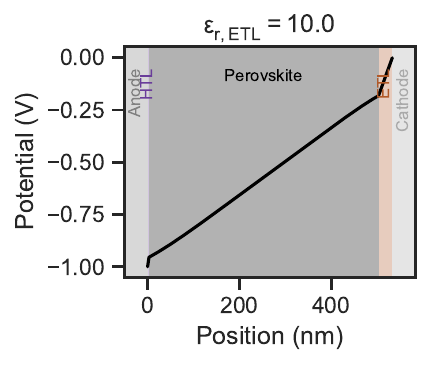}
    \end{subfigure}
    \begin{subfigure}[t]{0.49\textwidth}
        \captionsetup{justification=raggedright, singlelinecheck=false, position=top}
        \subcaption{}\label{p5:fs5c}
        \includegraphics[scale=0.9]{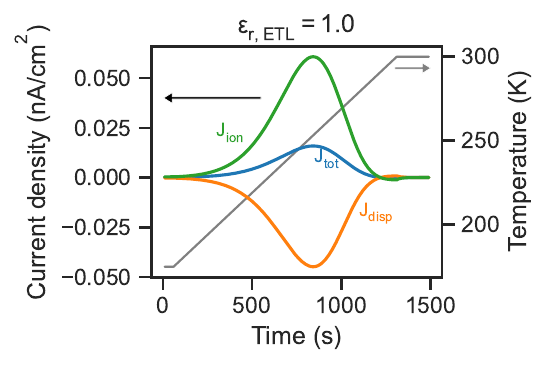}
    \end{subfigure}
    \begin{subfigure}[t]{0.49\textwidth}
        \captionsetup{justification=raggedright, singlelinecheck=false, position=top}
        \subcaption{}\label{p5:fs5d}
        \includegraphics[scale=0.9]{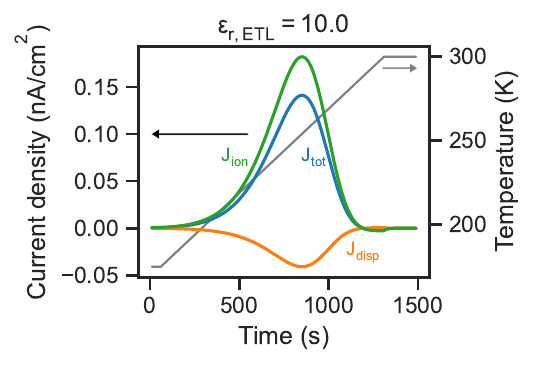}
    \end{subfigure}
    \caption{Simulated potential of a perovskite solar cell with a dielectric constant of  \subref{p5:fs5a} $\epsilon_{\mathrm{r,ETL}} = 1.0$ and \subref{p5:fs5b}  $\epsilon_{\mathrm{r,ETL}} = 10.0$ right after releasing the voltage pulse in a TAIC measurement. \subref{p5:fs5c} Current contributions of the ionic current $J_{\mathrm{ion}}$ and the displacement current $J_{\mathrm{disp}}$ to the total current $J_{\mathrm{tot}}$ during the TAIC measurement of the devices with dielectric constant of  $\epsilon_{\mathrm{r,ETL}} = 1.0$ and \subref{p5:fs5d}  $\epsilon_{\mathrm{r,ETL}} = 10.0$.   }
    \label{p5:fs5}
\end{figure}

The drift current of mobile ions to the perovskite/CTL interfaces depends on the electric field in the perovskite bulk according to Equation \ref{p5:equ1_ionicCurrent} in the main text. The electric field, in turn, depends on how much of the built-in potential of the perovskite solar cell drops over the perovskite. If organic CTLs are present, significant parts of the built-in potential can drop in the CTLs due to their low dielectric constant. Then, less potential drops over the perovskite, reducing the electric field in the perovskite bulk. This is illustrated in Figure \ref{p5:fs5a} and \subref{p5:fs5b}, which show drift-diffusion simulations of the potential of a perovskite solar cell immediately after removing the voltage pulse in a TAIC measurement. 
 \\
In Figure \ref{p5:fs5a}, a significantly higher fraction of the potential drops over the ETL compared to Figure \ref{p5:fs5b} due the lower dielectric constant in the ETL ($\epsilon_{\mathrm{r,ETL}} = 1$ and $\epsilon_{\mathrm{r,ETL}} = 10$, with $\epsilon_{\mathrm{r,Pero}} = 50$). This difference results in a lower electric field inside the perovskite. We can estimate the potential drop in the perovskite by considering the dielectric constants of the individual layers. We first assume that the electric displacement field $D$ throughout the device is constant:
\begin{equation}
    D = \epsilon_{\mathrm{0}} \epsilon_{\mathrm{r,HTL}} E_{\mathrm{HTL}} = \epsilon_{\mathrm{0}} \epsilon_{\mathrm{r,Pero}} E_{\mathrm{Pero}} = \epsilon_{\mathrm{0}} \epsilon_{\mathrm{r,ETL}} E_{\mathrm{ETL}}
\end{equation}
where $\epsilon_{\mathrm{0}}$ is the vacuum permittivity and $\epsilon_{\mathrm{r,HTL}}$, $\epsilon_{\mathrm{r,Pero}}$, and $\epsilon_{\mathrm{r,ETL}}$ are the relative dielectric constants of the different layers. $E_{\mathrm{HTL}}$,  $E_{\mathrm{Pero}}$, and $E_{\mathrm{ETL}}$ refer to the electric field in the individual layers. Next, we express the potential drops in the individual layers in terms of the electric field and the dielectric constant in the perovskite: 
\begin{align}
    \Delta V_{\mathrm{HTL}} &= E_{\mathrm{HTL}} d_{\mathrm{HTL}} = \frac{\epsilon_{\mathrm{r,Pero}} } {\epsilon_{\mathrm{r,HTL}}} d_{\mathrm{HTL}} E_{\mathrm{Pero}} \\
    \Delta V_{\mathrm{Pero}} &= E_{\mathrm{Pero}} d_{\mathrm{Pero}}  \label{p5:equ:PotDropPero}\\
    \Delta V_{\mathrm{ETL}} &= E_{\mathrm{ETL}} d_{\mathrm{ETL}} = \frac{\epsilon_{\mathrm{r,Pero}} } {\epsilon_{\mathrm{r,ETL}}} d_{\mathrm{ETL}} E_{\mathrm{Pero}} 
\end{align}
where $d_{\mathrm{HTL}}$,  $d_{\mathrm{Pero}}$, and $d_{\mathrm{ETL}}$ are the thicknesses of the different layers. The sum of the potential drops has to equal the built-in voltage $V_{\mathrm{bi}}$:
\begin{align}
    V_{\mathrm{bi}} &= \Delta V_{\mathrm{HTL}} + \Delta V_{\mathrm{Pero}} + \Delta V_{\mathrm{ETL}} \\
            &= E_{\mathrm{Pero}} \big( d_{\mathrm{Pero}} + \frac{\epsilon_{\mathrm{r,Pero}}}{\epsilon_{\mathrm{r,HTL}}}d_{\mathrm{HTL}} + \frac{\epsilon_{\mathrm{r,Pero}}}{\epsilon_{\mathrm{r,ETL}}}d_{\mathrm{ETL}}  \big)
\end{align}
We can rearrange this expression for the electric field in the perovskite: 
\begin{align}
    E_{\mathrm{Pero}}  &= V_{\mathrm{bi}} \big( d_{\mathrm{Pero}} + \frac{\epsilon_{\mathrm{r,Pero}}}{\epsilon_{\mathrm{r,HTL}}}d_{\mathrm{HTL}} + \frac{\epsilon_{\mathrm{r,Pero}}}{\epsilon_{\mathrm{r,ETL}}}d_{\mathrm{ETL}}  \big)^{-1}
\end{align}
With Equation \ref{p5:equ:PotDropPero}, we can approximate the potential drop in the perovskite:
\begin{equation}
    \Delta V_{\mathrm{Pero}}  = d_{\mathrm{Pero}}V_{\mathrm{bi}} \big( d_{\mathrm{Pero}} + \frac{\epsilon_{\mathrm{r,Pero}}}{\epsilon_{\mathrm{r,HTL}}}d_{\mathrm{HTL}} + \frac{\epsilon_{\mathrm{r,Pero}}}{\epsilon_{\mathrm{r,ETL}}}d_{\mathrm{ETL}}  \big)^{-1}
\end{equation}
We define the correction factor $b$ as the fraction of the built-in potential that drops within the perovskite: 
\begin{align}
    b &= \frac{\Delta V_{\mathrm{Pero}} }{V_{\mathrm{bi}}} \notag \\
      &= \big( 1 + \frac{\epsilon_{\mathrm{r,Pero}}d_{\mathrm{HTL}}}{\epsilon_{\mathrm{r,HTL}}d_{\mathrm{Pero}}} + \frac{\epsilon_{\mathrm{r,Pero}}d_{\mathrm{ETL}}}{\epsilon_{\mathrm{r,ETL}}d_{\mathrm{Pero}}}  \big)^{-1}
      \label{p5:equ:CorrectionFactor}
\end{align}
For the cases in Figure \ref{p5:fs5a} and \subref{p5:fs5b} and the device parameters in Table \ref{p5:tabS3:DDSimParameters} (without any doping in the ETL) we calculate a correction factor of \SI{0.25}{} and \SI{0.73}{}, respectively, meaning that the potential drop within the perovskite is approximately \SI{0.25}{V} and \SI{0.73}{V} ($V_{\mathrm{bi}}$ is \SI{1}{V}). This is in good agreement with the simulations. \\
Next to impacting the electric field in the perovskite bulk, potential drops within the CTLs also impact the current that is measured in TAIC and in general current transient measurements. This is illustrated in Figure \ref{p5:fs5c} and \subref{p5:fs5d} for the dielectric constants of $\epsilon_{\mathrm{r,ETL}} = 1$ in and $\epsilon_{\mathrm{r,ETL}} = 10$ for the ETL. The total extracted current in both cases consists of the ionic current and a displacement current in the perovskite, which are opposite in sign. For a lower dielectric constant of $\epsilon_{\mathrm{r,ETL}} = 1$ the displacement current makes up a significant part of the total current $J_{\mathrm{tot}}$, and the total current is only a fraction of the ionic current $J_{\mathrm{ion}}$. This occurs due to the higher potential drop and, consequently, the higher electric field in the ETL. Ions that accumulate at the interface lead to a change of the electric field and consequently a displacement current. For a higher electric field in the ETL, the relative change of the electric field due to accumulated ionic carriers becomes lower, limiting the displacement current in the ETL. This also leads to a displacement current in the perovskite bulk because the current has to be constant throughout the device. \\
When integrating the total current, the ion density is significantly underestimated. Here, it would lead to an estimated ion density of \SI{1.8e15}{cm^{-3}}, significantly lower than the actual value of \SI{1.0e16}{cm^{-3}}. We can, however, account for underestimation due to the displacement current with the correction factor in Equation \ref{p5:equ3_ionicDensity}, resulting in a more accurate ion density of \SI{7.2e15}{cm^{-3}}. \\
With Equation \ref{p5:equ:CorrectionFactor} and the parameters in Table \ref{p5:tabs2_CorrectionFactor}, we can estimate the correction factor for the \ce{MAPbI3} and the triple-cation device to be 0.76 and 0.66, respectively. 

\begin{table}[hbt]
\centering
\setlength{\tabcolsep}{2pt} 
\renewcommand{\arraystretch}{1.4} 
{\small
\begin{tabular}{c c c c }
Parameter & \ce{MAPbI3}  &  Triple-cation & Comment \\ \hline
Thickness HTL (nm) & 2 & 2 & Estimate \\
Thickness perovskite (nm) & 470.0 & 550.0 & Measured with profilometer \\
Thickness ETL (nm) & 20.0 & 30.0 & Estimate \\
Dielectric constant HTL & 4.0 & 4.0 & Estimate \\
Dielectric constant perovskite & 33.0 & 43.0 & From C-f measurements\\
Dielectric constant ETL & 5.0 & 5.0 & Estimate \\ \hline
\end{tabular}
}
\caption{Parameter values used to calculate the correction factor of the \ce{MAPbI3} and the triple-cation device. }
\label{p5:tabs2_CorrectionFactor}
\end{table} 

\section*{Additional information }
\begin{figure}[hbt]
\centering
    \begin{subfigure}[t]{0.48\textwidth}
        \captionsetup{justification=raggedright, singlelinecheck=false, position=top}
        \subcaption{}\label{p5:fs4a}
        \includegraphics[scale=0.9]{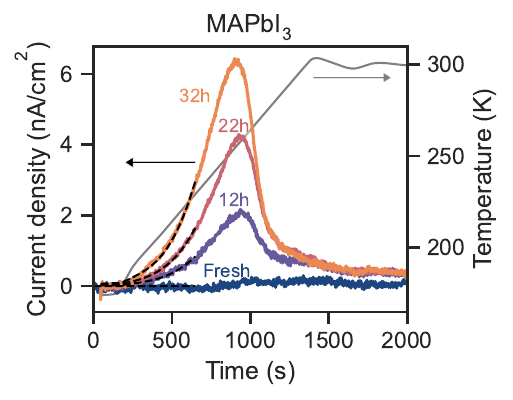}
    \end{subfigure}
    \begin{subfigure}[t]{0.48\textwidth}
        \captionsetup{justification=raggedright, singlelinecheck=false, position=top}
        \subcaption{}\label{p5:fs4b}
        \includegraphics[scale=0.9]{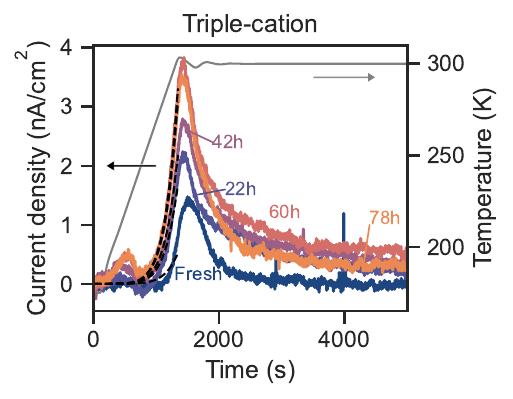}
    \end{subfigure}
    \caption{Thermally activated ion current measurements of a second \subref{p5:fs4a}  \ce{MAPbI3} and \subref{p5:fs4b} triple-cation perovskite solar cell for different stressing durations. The black dashed lines represent fits. The gray line represents an exemplary temperature sweep. The extracted ion parameters are shown in Table \ref{p5:tabS4_IonParametersSecondDevice}.}
    \label{p5:fs4}
\end{figure} 
\begin{table}[hbt]
\centering
\setlength{\tabcolsep}{8pt} 
\renewcommand{\arraystretch}{1.4} 
{\small
\begin{tabular}{c c c c c c  }
Device & Stressing & $E_{\mathrm{a}}$ (eV) & $\sigma_{\mathrm{ion,300K}}$ (\SI{}{S/cm}) &  $N_{\mathrm{ion}}$ (\SI{}{cm^{-3}}) & $D_{\mathrm{ion,300K}}$ (\SI{}{cm^2/s}) \\ \hline
\multirow{4}{*}{\ce{MAPbI_3}} & Fresh & \multirow{4}{*}{0.25} & - &  -  & -  \\
                & 12h &   & $9.0 \pm 0.7 \cdot 10^{-13} $ &  $5.8\pm0.2\cdot 10^{17}$  & $2.5 \pm 0.2 \cdot 10^{-13} $ \\
                & 22h &   & $1.8 \pm 0.1 \cdot 10^{-12} $ & $9.3\pm0.2\cdot 10^{17}$  & $3.1 \pm 0.2 \cdot 10^{-13} $  \\
                & 32h &   & $3.2\pm 0.2 \cdot 10^{-12} $ & $12.1\pm0.1\cdot 10^{17}$  & $4.2 \pm 0.3 \cdot 10^{-13} $  \\ \hline
\multirow{5}{*}{Triple-cation} & Fresh & \multirow{5}{*}{0.42} & $5.7 \pm 0.8 \cdot 10^{-14} $ & $2.9\pm0.3\cdot 10^{17}$ &  $3.1\pm 0.5 \cdot 10^{-14} $  \\
                & 22h &   & $1.8 \pm 0.2 \cdot 10^{-13} $ & $13.1\pm0.4\cdot 10^{17}$ &  $2.2 \pm 0.3 \cdot 10^{-14} $ \\
                & 42h &   & $2.4 \pm 0.3 \cdot 10^{-13} $ & $19.1\pm0.5\cdot 10^{17}$ &  $2.0 \pm0.3 \cdot 10^{-14} $\\
                & 60h &   & $3.6 \pm 0.5 \cdot 10^{-13} $ & $27.1\pm0.5\cdot 10^{17}$ & $2.1 \pm 0.3 \cdot 10^{-14} $ \\
                & 78h &   & $3.4 \pm 0.4 \cdot 10^{-13} $ & $24.2\pm0.5\cdot 10^{17}$ &   $2.2\pm 0.3 \cdot 10^{-14} $ \\ \hline
\end{tabular}
}
\caption{Estimated values of the activation energy $E_{\mathrm{a}}$, ionic conductivity at \SI{300}{K} $\sigma_{\mathrm{ion,300K}}$, ion density $N_{\mathrm{ion}}$, and diffusion coefficient at \SI{300}{K} $D_{\mathrm{ion,300K}}$ for the second \ce{MAPbI3} and the triple-cation devices in Figure \ref{p5:fs4}. The values were extracted from the low-temperature fit and the integral of the TAIC measurements. The error of $N_{\mathrm{ion}}$ is estimated from the minimum detectable ion density based on the noise of the current and the diffusion coefficient at the temperature of the current peaks. The errors of the $\sigma_{\mathrm{ion,300K}}$ correspond to the fitting error. The error of $D_{\mathrm{ion,300K}}$ is propagated based on the errors of $N_{\mathrm{ion}}$ and $\sigma_{\mathrm{ion,300K}}$. For the fresh \ce{MAPbI3} device, we could not extract any values because the current is below the noise.}
\label{p5:tabS4_IonParametersSecondDevice}
\end{table} 

\begin{figure}[hbt]
\centering
    \begin{subfigure}[t]{0.49\textwidth}
        \captionsetup{justification=raggedright, singlelinecheck=false, position=top}
        \subcaption{}\label{p5:fs6a}
        \includegraphics[scale=0.9]{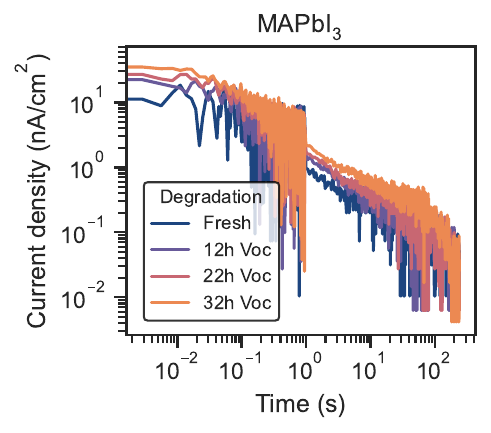}
    \end{subfigure}
    \begin{subfigure}[t]{0.49\textwidth}
        \captionsetup{justification=raggedright, singlelinecheck=false, position=top}
        \subcaption{}\label{p5:fs6b}
        \includegraphics[scale=0.9]{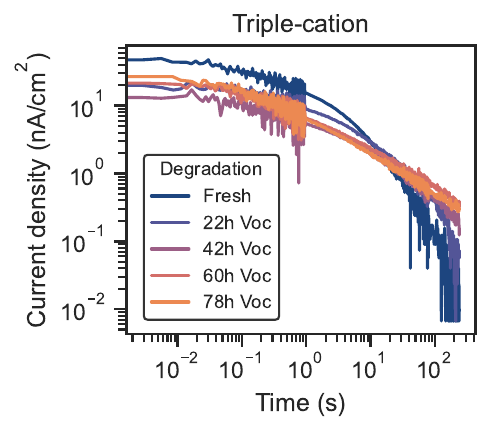}
    \end{subfigure}
    \caption{Current transient measurements after switching off the applied bias after cooling the devices down to \SI{175}{K} in a TAIC measurement of \subref{p5:fs6a} a \ce{MAPbI3} device and \subref{p5:fs6b} a triple-cation device. The noise changes around \SI{1}{s} because the integration time of the source-measure unit changes. }
    \label{p5:fs6}
\end{figure}

\begin{table}[hbt]
\centering
\setlength{\tabcolsep}{8pt} 
\renewcommand{\arraystretch}{1.4} 
{\small
\begin{tabular}{c c c  }
Device & Stressing &  $N_{\mathrm{ion}}D_{\mathrm{0,ion}} \mathrm{(cm\,s)^{-1}}$ \\ \hline
\multirow{4}{*}{\ce{MAPbI_3}} & Fresh &  $2.7 \pm 0.2 \cdot 10^{9} $  \\
                & 12h &    $1.8 \pm 0.1 \cdot 10^{10} $ \\
                & 22h &    $2.7 \pm 0.2 \cdot 10^{10} $  \\
                & 32h &    $3.5 \pm 0.2 \cdot 10^{10} $  \\ \hline
\multirow{5}{*}{Triple-cation} & Fresh &  $1.5 \pm 0.1 \cdot 10^{9} $  \\
                & 22h &    $13.8 \pm 0. 5\cdot 10^{9} $ \\
                & 42h &    $20.9 \pm 0.2 \cdot 10^{9} $\\
                & 60h &    $32.1 \pm 0.5 \cdot 10^{9} $ \\
                & 78h &      $43.8 \pm 0.8 \cdot 10^{9} $ \\ \hline
\end{tabular}
}
\caption{Fitting value of the product of ion density and diffusion coefficient prefactor $N_{\mathrm{ion}}D_{\mathrm{0,ion}}$ extracted from the low temperature fits in Figure \ref{p5:f3}. The errors correspond to the fitting error.}
\label{p5:tabS3:MeasurementDevice1FittingValues}
\end{table}

\begin{table}[h!]
\centering
\setlength{\tabcolsep}{3pt} 
\renewcommand{\arraystretch}{1.2} 
{\footnotesize
\begin{tabular}{c c c c c c c c c }
 & Set $E_{\mathrm{a}}$  &  Fit $E_{\mathrm{a}}$ & Set $\sigma_{\mathrm{0,ion}}$  &  Fit $\sigma_{\mathrm{0,ion}}$  & Set $N_{\mathrm{ion}}$ & Approx. $N_{\mathrm{ion}}$ & Set $\mu_{\mathrm{0,ion}}$ & Approx. $\mu_{\mathrm{0,ion}}$  \\
  & (eV) &  (eV) & $\mathrm{(S/cm)}$ &  $\mathrm{(S/cm)}$ &  $\mathrm{(1/cm^3)}$ & $\mathrm{(1/cm^3)}$ & $\mathrm{(cm^2/Vs)}$ & $\mathrm{(cm^2/Vs)}$  \\ \hline
\multirow{8}{*}{Ion-limited} & \multirow{4}{*}{0.3} & \multirow{4}{*}{0.29} & $1.1 \cdot 10^{-9}$ & $7.3\cdot 10^{-10}$ & $10^{15}$ & $1.2 \cdot 10^{15}$ & \multirow{4}{*}{$7 \cdot 10^{-6}$}  &  $3.9 \cdot 10^{-6}$ \\ 
 &  &  & $2.4 \cdot 10^{-9}$ & $1.6\cdot 10^{-9}$ &  $2.2 \cdot 10^{15}$ & $2.5 \cdot 10^{15}$ &  &  $4.1 \cdot 10^{-6}$ \\ 
  &  &  & $5.2 \cdot 10^{-9}$ & $3.6\cdot 10^{-9}$ & $4.6 \cdot 10^{15}$ & $5.2 \cdot 10^{15}$ &  &  $4.4 \cdot 10^{-6}$ \\ 
   &  &  & $1.1 \cdot 10^{-8}$ & $7.8\cdot 10^{-9}$ & $10^{16}$ & $1.0 \cdot 10^{16}$ &  &  $4.8 \cdot 10^{-6}$ \\ \cline{2-9}
 & \multirow{4}{*}{0.6} & \multirow{4}{*}{0.61} & $4.8 \cdot 10^{-6}$ & $6.5\cdot 10^{-6}$ & $10^{15}$ & $1.2 \cdot 10^{15}$ & \multirow{4}{*}{$3.0 \cdot 10^{-2}$}  &  $3.4 \cdot 10^{-2}$ \\ 
 &  &  & $1.0 \cdot 10^{-5}$ & $1.6\cdot 10^{-5}$ & $2.2 \cdot 10^{15}$ & $2.5 \cdot 10^{15}$ &  &  $4.1 \cdot 10^{-2}$ \\ 
  &  &  & $2.2 \cdot 10^{-5}$ & $3.8\cdot 10^{-5}$ & $4.6 \cdot 10^{15}$ & $5.2 \cdot 10^{15}$ &  &  $4.5 \cdot 10^{-2}$ \\ 
   &  &  & $4.8 \cdot 10^{-5}$ & $8.3\cdot 10^{-5}$ & $10^{16}$ & $1.0 \cdot 10^{16}$ &  &  $5.1 \cdot 10^{-2}$ \\ \hline
\multirow{8}{*}{Field-limited} & \multirow{4}{*}{0.3} & \multirow{4}{*}{0.28} & $1.1 \cdot 10^{-8}$ & $4.9\cdot 10^{-9}$ & $10^{17}$ & $2.6 \cdot 10^{16}$ & \multirow{4}{*}{$7.0 \cdot 10^{-7}$}  &  - \\ 
 &  &  & $2.4 \cdot 10^{-8}$ & $1.0\cdot 10^{-8}$ & $2.2 \cdot 10^{17}$ & $3.4 \cdot 10^{16}$ &  &  - \\ 
  &  &  & $5.2 \cdot 10^{-8}$ & $2.1 \cdot 10^{-8}$ & $4.6 \cdot 10^{17}$ & $3.8 \cdot 10^{16}$ &  &  - \\ 
   &  &  & $1.1 \cdot 10^{-7}$ & $4.4\cdot 10^{-8}$ & $10^{18}$ & $4.0 \cdot 10^{16}$ &  &  - \\ \cline{2-9}
 & \multirow{4}{*}{0.6} & \multirow{4}{*}{0.56} & $4.8 \cdot 10^{-5}$ & $1.1\cdot 10^{-5}$ & $10^{17}$ & $3.1 \cdot 10^{16}$ & \multirow{4}{*}{$3.0 \cdot 10^{-3}$}  &  - \\ 
 &  &  & $1.0 \cdot 10^{-4}$ & $2.5\cdot 10^{-5}$ & $2.2 \cdot 10^{17}$ & $3.5 \cdot 10^{16}$ &  &  - \\ 
  &  &  & $2.2 \cdot 10^{-4}$ & $5.3\cdot 10^{-5}$ & $4.6 \cdot 10^{17}$ & $3.8 \cdot 10^{16}$ &  &  - \\ 
   &  &  & $4.8 \cdot 10^{-4}$ & $1.1\cdot 10^{-4}$ & $10^{18}$ & $4.0 \cdot 10^{16}$ &  &  - \\ 
\end{tabular}
}
\caption{Values for the activation energy $E_{\mathrm{a}}$, ion density $N_{\mathrm{ion}}$, and mobility $\mu_{\mathrm{0,ion}}$ extracted from the drift-diffusion simulations in Figure \ref{p5:fs1}. The values were extracted by fitting the low-temperature current and integrating the total current of the TAIC measurements. Because the ion density is significantly underestimated in the field-limited case, we do not determine the ionic mobility.}
\label{p5:tabs1_FittingDD}
\end{table} 

\begin{table}[hbt]
\centering
\setlength{\tabcolsep}{8pt} 
\renewcommand{\arraystretch}{1.4} 
{\small
\begin{tabular}{c c c  }
Device & Stressing &  $N_{\mathrm{ion}} \mathrm{(cm^{-3}})$ \\ \hline
\multirow{4}{*}{\ce{MAPbI_3}} & Fresh &  $9.8 \cdot 10^{15} $  \\
                & 12h &    $1.2 \cdot 10^{16} $ \\
                & 22h &    $1.3 \cdot 10^{16} $  \\
                & 32h &    $1.8 \cdot 10^{16} $  \\ \hline
\multirow{5}{*}{Triple-cation} & Fresh &  $4.8 \cdot 10^{16} $ \\
                & 22h &    $5.5 \cdot 10^{16} $ \\
                & 42h &    $5.1 \cdot 10^{16} $\\
                & 60h &    $6.8 \cdot 10^{16} $ \\
                & 78h &    $5.7 \cdot 10^{16} $ \\ \hline
\end{tabular}
}
\caption{Ion density estimated from current transient measurements measured at \SI{175}{K} in Figure \ref{p5:fs6}.}
\label{p5:tabS5:Nion_LowTtransients}
\end{table}

\begin{figure}[hbt]
\centering
    \begin{subfigure}[t]{0.49\textwidth}
        \captionsetup{justification=raggedright, singlelinecheck=false, position=top}
        \subcaption{}\label{p5:fs2a}
        \includegraphics[scale=0.9]{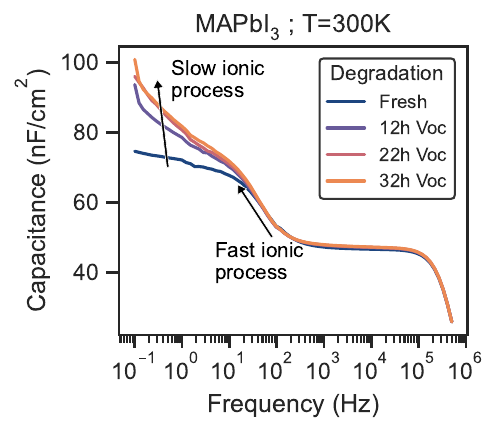}
    \end{subfigure}
    \begin{subfigure}[t]{0.49\textwidth}
        \captionsetup{justification=raggedright, singlelinecheck=false, position=top}
        \subcaption{}\label{p5:fs2b}
        \includegraphics[scale=0.9]{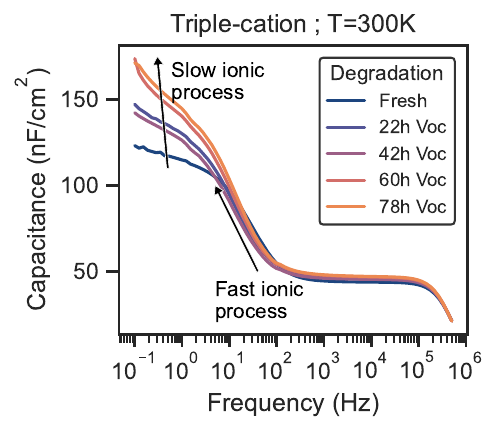}
    \end{subfigure}
    \begin{subfigure}[t]{0.49\textwidth}
        \captionsetup{justification=raggedright, singlelinecheck=false, position=top}
        \subcaption{}\label{p5:fs2c}
        \includegraphics[scale=0.9]{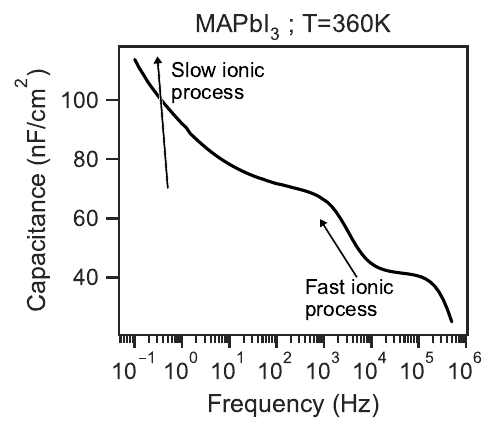}
    \end{subfigure}
    \begin{subfigure}[t]{0.49\textwidth}
        \captionsetup{justification=raggedright, singlelinecheck=false, position=top}
        \subcaption{}\label{p5:fs2d}
        \includegraphics[scale=0.9]{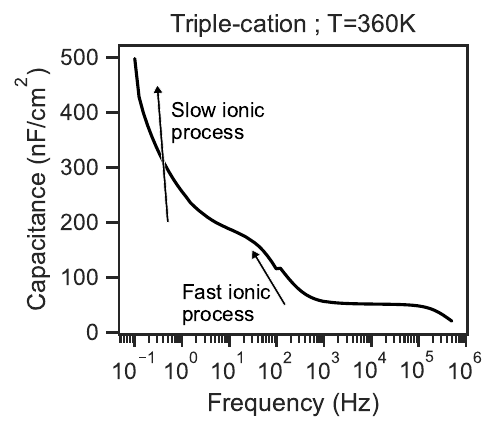}
    \end{subfigure}
    \caption{Capacitance frequency measurements of \subref{p5:fs2a} a \ce{MAPbI3} device at \SI{300}{K} after different stressing durations at $V_{\mathrm{oc}}$,  \subref{p5:fs2b} a triple-cation device at \SI{300}{K} after different stressing durations at $V_{\mathrm{oc}}$, \subref{p5:fs2c} a \ce{MAPbI3} device at \SI{360}{K} and \subref{p5:fs2d} a triple-cation device at \SI{360}{K}. }
    \label{p5:fs2}
\end{figure} 

\begin{figure}[hbt]
\centering
    \begin{subfigure}[t]{0.49\textwidth}
        \captionsetup{justification=raggedright, singlelinecheck=false, position=top}
        \subcaption{}\label{p5:fs3a}
        \includegraphics[scale=0.82]{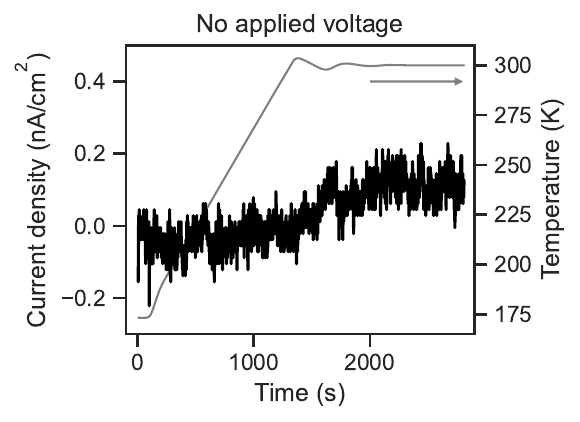}
    \end{subfigure}
    \begin{subfigure}[t]{0.49\textwidth}
        \captionsetup{justification=raggedright, singlelinecheck=false, position=top}
        \subcaption{}\label{p5:fs3b}
        \includegraphics[scale=0.82]{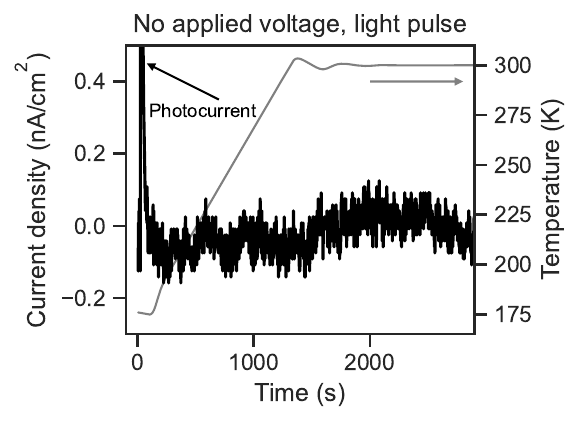}
    \end{subfigure}
    \caption{TAIC measurement of a triple-cation device \subref{p5:fs3a} without any applied voltage during the cool-down and \subref{p5:fs3b} without any applied voltage during the cool-down and briefly illuminated at \SI{175}{K}. }
    \label{p5:fs3}
\end{figure} 

\begin{figure}[hbt]
\centering
    \begin{subfigure}[t]{0.49\textwidth}
        \captionsetup{justification=raggedright, singlelinecheck=false, position=top}
        \subcaption{}\label{p5:fs8a}
        \includegraphics[scale=0.9]{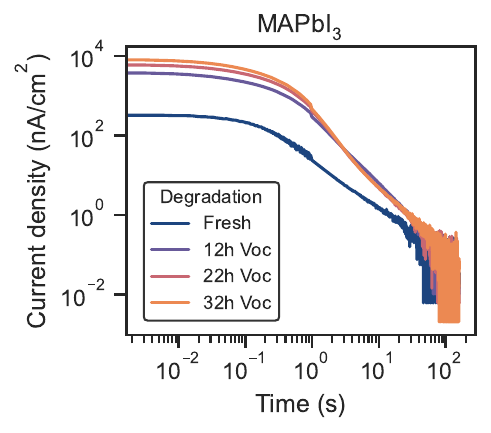}
    \end{subfigure}
    \begin{subfigure}[t]{0.49\textwidth}
        \captionsetup{justification=raggedright, singlelinecheck=false, position=top}
        \subcaption{}\label{p5:fs8b}
        \includegraphics[scale=0.9]{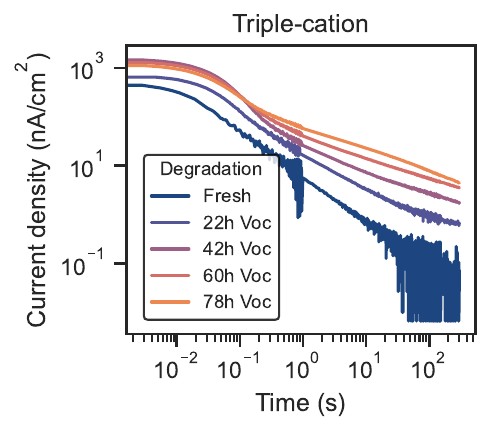}
    \end{subfigure}
    \caption{Current transient measurements at \SI{300}{K} of \subref{p5:fs8a} a \ce{MAPbI3} device and \subref{p5:fs8b} a triple-cation device. The noise changes around \SI{1}{s} because the integration time of the source-measure unit changes. To avoid the devices recovering before starting the TAIC measurements, we limit the measurement time of the transient current measurements. }
    \label{p5:fs8}
\end{figure}

\begin{table}[hbt]
\centering
\setlength{\tabcolsep}{8pt} 
\renewcommand{\arraystretch}{1.4} 
{\small
\begin{tabular}{c c c  }
Device & Stressing &  $N_{\mathrm{ion}} \mathrm{(cm^{-3}})$ \\ \hline
\multirow{4}{*}{\ce{MAPbI_3}} & Fresh &  $5.8 \cdot 10^{16} $  \\
                & 12h &    $5.2 \cdot 10^{17} $ \\
                & 22h &    $7.4 \cdot 10^{17} $  \\
                & 32h &    $9.1 \cdot 10^{17} $  \\ \hline
\multirow{5}{*}{Triple-cation} & Fresh &  $2.5 \cdot 10^{16} $ \\
                & 22h &    $1.4 \cdot 10^{17} $ \\
                & 42h &    $3.4 \cdot 10^{17} $\\
                & 60h &    $6.3 \cdot 10^{17} $ \\
                & 78h &    $9.0 \cdot 10^{17} $ \\ \hline
\end{tabular}
}
\caption{Ion density estimated from current transient measurements measured at \SI{300}{K} in Figure \ref{p5:fs8}.}
\label{p5:tabS8:Nion_300Ktransients}
\end{table}

\begin{figure}[hbt]
\centering
    \begin{subfigure}[t]{0.49\textwidth}
        \captionsetup{justification=raggedright, singlelinecheck=false, position=top}
        \subcaption{}\label{p5:fs7a}
        \includegraphics[scale=0.9]{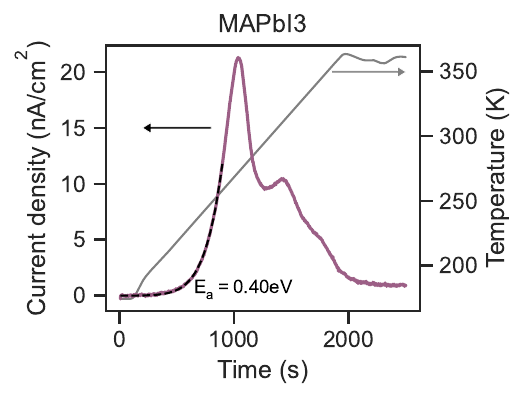}
    \end{subfigure}
    \begin{subfigure}[t]{0.49\textwidth}
        \captionsetup{justification=raggedright, singlelinecheck=false, position=top}
        \subcaption{}\label{p5:fs7b}
        \includegraphics[scale=0.9]{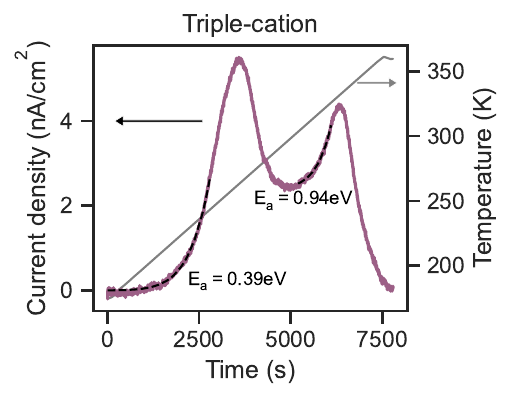}
    \end{subfigure}
    \caption{TAIC measurements starting and ending at \SI{360}{K} of \subref{p5:fs7a} a \ce{MAPbI3} and \subref{p5:fs7b} a triple-cation device. The temperature sweep speed for the \ce{MAPbI3} device is \SI{0.1}{K/s} and for the triple-cation device \SI{0.025}{K/s}. Dashed lines indicate fits.}
    \label{p5:fs7}
\end{figure} 

\clearpage
\bibliographystyle{achemso}
\bibliography{SI}

\providecommand{\latin}[1]{#1}
\makeatletter
\providecommand{\doi}
  {\begingroup\let\do\@makeother\dospecials
  \catcode`\{=1 \catcode`\}=2 \doi@aux}
\providecommand{\doi@aux}[1]{\endgroup\texttt{#1}}
\makeatother
\providecommand*\mcitethebibliography{\thebibliography}
\csname @ifundefined\endcsname{endmcitethebibliography}  {\let\endmcitethebibliography\endthebibliography}{}
\begin{mcitethebibliography}{1}
\providecommand*\natexlab[1]{#1}
\providecommand*\mciteSetBstSublistMode[1]{}
\providecommand*\mciteSetBstMaxWidthForm[2]{}
\providecommand*\mciteBstWouldAddEndPuncttrue
  {\def\EndOfBibitem{\unskip.}}
\providecommand*\mciteBstWouldAddEndPunctfalse
  {\let\EndOfBibitem\relax}
\providecommand*\mciteSetBstMidEndSepPunct[3]{}
\providecommand*\mciteSetBstSublistLabelBeginEnd[3]{}
\providecommand*\EndOfBibitem{}
\mciteSetBstSublistMode{f}
\mciteSetBstMaxWidthForm{subitem}{(\alph{mcitesubitemcount})}
\mciteSetBstSublistLabelBeginEnd
  {\mcitemaxwidthsubitemform\space}
  {\relax}
  {\relax}

\bibitem[Futscher and Milić(2021)Futscher, and Milić]{futscher_mixed_2021}
Futscher,~M.~H.; Milić,~J.~V. Mixed {Conductivity} of {Hybrid} {Halide} {Perovskites}: {Emerging} {Opportunities} and {Challenges}. \emph{Frontiers in Energy Research} \textbf{2021}, \emph{9}, 629074\relax
\mciteBstWouldAddEndPuncttrue
\mciteSetBstMidEndSepPunct{\mcitedefaultmidpunct}
{\mcitedefaultendpunct}{\mcitedefaultseppunct}\relax
\EndOfBibitem
\end{mcitethebibliography}


\providecommand{\latin}[1]{#1}
\makeatletter
\providecommand{\doi}
  {\begingroup\let\do\@makeother\dospecials
  \catcode`\{=1 \catcode`\}=2 \doi@aux}
\providecommand{\doi@aux}[1]{\endgroup\texttt{#1}}
\makeatother
\providecommand*\mcitethebibliography{\thebibliography}
\csname @ifundefined\endcsname{endmcitethebibliography}  {\let\endmcitethebibliography\endthebibliography}{}
\begin{mcitethebibliography}{45}
\providecommand*\natexlab[1]{#1}
\providecommand*\mciteSetBstSublistMode[1]{}
\providecommand*\mciteSetBstMaxWidthForm[2]{}
\providecommand*\mciteBstWouldAddEndPuncttrue
  {\def\EndOfBibitem{\unskip.}}
\providecommand*\mciteBstWouldAddEndPunctfalse
  {\let\EndOfBibitem\relax}
\providecommand*\mciteSetBstMidEndSepPunct[3]{}
\providecommand*\mciteSetBstSublistLabelBeginEnd[3]{}
\providecommand*\EndOfBibitem{}
\mciteSetBstSublistMode{f}
\mciteSetBstMaxWidthForm{subitem}{(\alph{mcitesubitemcount})}
\mciteSetBstSublistLabelBeginEnd
  {\mcitemaxwidthsubitemform\space}
  {\relax}
  {\relax}

\bibitem[Thiesbrummel \latin{et~al.}(2024)Thiesbrummel, Shah, Gutierrez-Partida, Zu, Peña-Camargo, Zeiske, Diekmann, Ye, Peters, Brinkmann, Caprioglio, Dasgupta, Seo, Adeleye, Warby, Jeangros, Lang, Zhang, Albrecht, Riedl, Armin, Neher, Koch, Wu, Le~Corre, Snaith, and Stolterfoht]{thiesbrummel_ion-induced_2024}
Thiesbrummel,~J. \latin{et~al.}  Ion-induced field screening as a dominant factor in perovskite solar cell operational stability. \emph{Nature Energy} \textbf{2024}, \emph{9}, 664--676\relax
\mciteBstWouldAddEndPuncttrue
\mciteSetBstMidEndSepPunct{\mcitedefaultmidpunct}
{\mcitedefaultendpunct}{\mcitedefaultseppunct}\relax
\EndOfBibitem
\bibitem[Thiesbrummel \latin{et~al.}(2021)Thiesbrummel, Le~Corre, Peña‐Camargo, Perdigón‐Toro, Lang, Yang, Grischek, Gutierrez‐Partida, Warby, Farrar, Mahesh, Caprioglio, Albrecht, Neher, Snaith, and Stolterfoht]{thiesbrummel_universal_2021}
Thiesbrummel,~J. \latin{et~al.}  Universal {Current} {Losses} in {Perovskite} {Solar} {Cells} {Due} to {Mobile} {Ions}. \emph{Advanced Energy Materials} \textbf{2021}, \emph{11}, 2101447\relax
\mciteBstWouldAddEndPuncttrue
\mciteSetBstMidEndSepPunct{\mcitedefaultmidpunct}
{\mcitedefaultendpunct}{\mcitedefaultseppunct}\relax
\EndOfBibitem
\bibitem[Hart \latin{et~al.}(2024)Hart, Angus, Li, Khaleed, Calado, Durrant, Djurišić, Docampo, and Barnes]{hart_more_2024}
Hart,~L. J.~F.; Angus,~F.~J.; Li,~Y.; Khaleed,~A.; Calado,~P.; Durrant,~J.~R.; Djurišić,~A.~B.; Docampo,~P.; Barnes,~P. R.~F. More is different: mobile ions improve the design tolerances of perovskite solar cells. \emph{Energy \& Environmental Science} \textbf{2024}, \emph{17}, 7107--7118\relax
\mciteBstWouldAddEndPuncttrue
\mciteSetBstMidEndSepPunct{\mcitedefaultmidpunct}
{\mcitedefaultendpunct}{\mcitedefaultseppunct}\relax
\EndOfBibitem
\bibitem[Diethelm \latin{et~al.}(2025)Diethelm, Lukas, Smith, Dasgupta, Caprioglio, Futscher, Hany, and Snaith]{diethelm_probing_2025}
Diethelm,~M.; Lukas,~T.; Smith,~J.; Dasgupta,~A.; Caprioglio,~P.; Futscher,~M.; Hany,~R.; Snaith,~H.~J. Probing ionic conductivity and electric field screening in perovskite solar cells: a novel exploration through ion drift currents. \emph{Energy \& Environmental Science} \textbf{2025}, \emph{18}, 1385--1397\relax
\mciteBstWouldAddEndPuncttrue
\mciteSetBstMidEndSepPunct{\mcitedefaultmidpunct}
{\mcitedefaultendpunct}{\mcitedefaultseppunct}\relax
\EndOfBibitem
\bibitem[Diekmann \latin{et~al.}(2023)Diekmann, Peña-Camargo, Tokmoldin, Thiesbrummel, Warby, Gutierrez-Partida, Shah, Neher, and Stolterfoht]{diekmann_determination_2023}
Diekmann,~J.; Peña-Camargo,~F.; Tokmoldin,~N.; Thiesbrummel,~J.; Warby,~J.; Gutierrez-Partida,~E.; Shah,~S.; Neher,~D.; Stolterfoht,~M. Determination of {Mobile} {Ion} {Densities} in {Halide} {Perovskites} via {Low}-{Frequency} {Capacitance} and {Charge} {Extraction} {Techniques}. \emph{The Journal of Physical Chemistry Letters} \textbf{2023}, 4200--4210\relax
\mciteBstWouldAddEndPuncttrue
\mciteSetBstMidEndSepPunct{\mcitedefaultmidpunct}
{\mcitedefaultendpunct}{\mcitedefaultseppunct}\relax
\EndOfBibitem
\bibitem[Schmidt \latin{et~al.}(2023)Schmidt, Gutierrez-Partida, Stolterfoht, and Ehrler]{schmidt_impact_2023}
Schmidt,~M.~C.; Gutierrez-Partida,~E.; Stolterfoht,~M.; Ehrler,~B. Impact of {Mobile} {Ions} on {Transient} {Capacitance} {Measurements} of {Perovskite} {Solar} {Cells}. \emph{PRX Energy} \textbf{2023}, \emph{2}, 043011\relax
\mciteBstWouldAddEndPuncttrue
\mciteSetBstMidEndSepPunct{\mcitedefaultmidpunct}
{\mcitedefaultendpunct}{\mcitedefaultseppunct}\relax
\EndOfBibitem
\bibitem[Wang \latin{et~al.}(2018)Wang, Kaienburg, Klingebiel, Schillings, and Kirchartz]{wang_understanding_2018}
Wang,~S.; Kaienburg,~P.; Klingebiel,~B.; Schillings,~D.; Kirchartz,~T. Understanding {Thermal} {Admittance} {Spectroscopy} in {Low}-{Mobility} {Semiconductors}. \emph{The Journal of Physical Chemistry C} \textbf{2018}, \emph{122}, 9795--9803\relax
\mciteBstWouldAddEndPuncttrue
\mciteSetBstMidEndSepPunct{\mcitedefaultmidpunct}
{\mcitedefaultendpunct}{\mcitedefaultseppunct}\relax
\EndOfBibitem
\bibitem[Lang(1974)]{lang_deep-level_1974}
Lang,~D.~V. Deep-level transient spectroscopy: {A} new method to characterize traps in semiconductors. \emph{Journal of Applied Physics} \textbf{1974}, \emph{45}, 3023--3032\relax
\mciteBstWouldAddEndPuncttrue
\mciteSetBstMidEndSepPunct{\mcitedefaultmidpunct}
{\mcitedefaultendpunct}{\mcitedefaultseppunct}\relax
\EndOfBibitem
\bibitem[Awni \latin{et~al.}(2020)Awni, Song, Chen, Li, Wang, Razooqi, Chen, Wang, Ellingson, Li, and Yan]{awni_influence_2020}
Awni,~R.~A.; Song,~Z.; Chen,~C.; Li,~C.; Wang,~C.; Razooqi,~M.~A.; Chen,~L.; Wang,~X.; Ellingson,~R.~J.; Li,~J.~V.; Yan,~Y. Influence of {Charge} {Transport} {Layers} on {Capacitance} {Measured} in {Halide} {Perovskite} {Solar} {Cells}. \emph{Joule} \textbf{2020}, \emph{4}, 644--657\relax
\mciteBstWouldAddEndPuncttrue
\mciteSetBstMidEndSepPunct{\mcitedefaultmidpunct}
{\mcitedefaultendpunct}{\mcitedefaultseppunct}\relax
\EndOfBibitem
\bibitem[Reichert \latin{et~al.}(2020)Reichert, An, Woo, Walsh, Vaynzof, and Deibel]{reichert_probing_2020}
Reichert,~S.; An,~Q.; Woo,~Y.-W.; Walsh,~A.; Vaynzof,~Y.; Deibel,~C. Probing the ionic defect landscape in halide perovskite solar cells. \emph{Nature Communications} \textbf{2020}, \emph{11}, 6098\relax
\mciteBstWouldAddEndPuncttrue
\mciteSetBstMidEndSepPunct{\mcitedefaultmidpunct}
{\mcitedefaultendpunct}{\mcitedefaultseppunct}\relax
\EndOfBibitem
\bibitem[Futscher \latin{et~al.}(2020)Futscher, Gangishetty, Congreve, and Ehrler]{futscher_quantifying_2020}
Futscher,~M.~H.; Gangishetty,~M.~K.; Congreve,~D.~N.; Ehrler,~B. Quantifying mobile ions and electronic defects in perovskite-based devices with temperature-dependent capacitance measurements: {Frequency} vs time domain. \emph{The Journal of Chemical Physics} \textbf{2020}, \emph{152}, 044202\relax
\mciteBstWouldAddEndPuncttrue
\mciteSetBstMidEndSepPunct{\mcitedefaultmidpunct}
{\mcitedefaultendpunct}{\mcitedefaultseppunct}\relax
\EndOfBibitem
\bibitem[McGovern \latin{et~al.}(2021)McGovern, Koschany, Grimaldi, Muscarella, and Ehrler]{mcgovern_grain_2021}
McGovern,~L.; Koschany,~I.; Grimaldi,~G.; Muscarella,~L.~A.; Ehrler,~B. Grain {Size} {Influences} {Activation} {Energy} and {Migration} {Pathways} in {MAPbBr} $_{\textrm{3}}$ {Perovskite} {Solar} {Cells}. \emph{The Journal of Physical Chemistry Letters} \textbf{2021}, \emph{12}, 2423--2428\relax
\mciteBstWouldAddEndPuncttrue
\mciteSetBstMidEndSepPunct{\mcitedefaultmidpunct}
{\mcitedefaultendpunct}{\mcitedefaultseppunct}\relax
\EndOfBibitem
\bibitem[Schmidt and Ehrler(2025)Schmidt, and Ehrler]{schmidt_how_2025}
Schmidt,~M.~C.; Ehrler,~B. How {Many} {Mobile} {Ions} {Can} {Electrical} {Measurements} {Detect} in {Perovskite} {Solar} {Cells}? \emph{ACS Energy Letters} \textbf{2025}, 2457--2460\relax
\mciteBstWouldAddEndPuncttrue
\mciteSetBstMidEndSepPunct{\mcitedefaultmidpunct}
{\mcitedefaultendpunct}{\mcitedefaultseppunct}\relax
\EndOfBibitem
\bibitem[Baumann \latin{et~al.}(2015)Baumann, Väth, Rieder, Heiber, Tvingstedt, and Dyakonov]{baumann_identification_2015}
Baumann,~A.; Väth,~S.; Rieder,~P.; Heiber,~M.~C.; Tvingstedt,~K.; Dyakonov,~V. Identification of {Trap} {States} in {Perovskite} {Solar} {Cells}. \emph{The Journal of Physical Chemistry Letters} \textbf{2015}, \emph{6}, 2350--2354\relax
\mciteBstWouldAddEndPuncttrue
\mciteSetBstMidEndSepPunct{\mcitedefaultmidpunct}
{\mcitedefaultendpunct}{\mcitedefaultseppunct}\relax
\EndOfBibitem
\bibitem[Khan \latin{et~al.}(2022)Khan, Schwenzer, Lehr, Paetzold, and Lemmer]{khan_emergence_2022}
Khan,~M.~R.; Schwenzer,~J.~A.; Lehr,~J.; Paetzold,~U.~W.; Lemmer,~U. Emergence of {Deep} {Traps} in {Long}-{Term} {Thermally} {Stressed} {CH}$_{\textrm{3}}$ {NH}$_{\textrm{3}}$ {PbI}$_{\textrm{3}}$ {Perovskite} {Revealed} by {Thermally} {Stimulated} {Currents}. \emph{The Journal of Physical Chemistry Letters} \textbf{2022}, \emph{13}, 552--558\relax
\mciteBstWouldAddEndPuncttrue
\mciteSetBstMidEndSepPunct{\mcitedefaultmidpunct}
{\mcitedefaultendpunct}{\mcitedefaultseppunct}\relax
\EndOfBibitem
\bibitem[Gordillo \latin{et~al.}(2017)Gordillo, Otálora, and Reinoso]{gordillo_trap_2017}
Gordillo,~G.; Otálora,~C.~A.; Reinoso,~M.~A. Trap center study in hybrid organic-inorganic perovskite using thermally stimulated current ({TSC}) analysis. \emph{Journal of Applied Physics} \textbf{2017}, \emph{122}, 075304\relax
\mciteBstWouldAddEndPuncttrue
\mciteSetBstMidEndSepPunct{\mcitedefaultmidpunct}
{\mcitedefaultendpunct}{\mcitedefaultseppunct}\relax
\EndOfBibitem
\bibitem[Xu \latin{et~al.}(2020)Xu, Wang, Han, Kamata, and Ma]{xu_suppression_2020}
Xu,~Z.; Wang,~L.; Han,~Q.; Kamata,~Y.; Ma,~T. Suppression of {Iodide} {Ion} {Migration} via {Sb}$_{\textrm{2}}$ {S}$_{\textrm{3}}$ {Interfacial} {Modification} for {Stable} {Inorganic} {Perovskite} {Solar} {Cells}. \emph{ACS Applied Materials \& Interfaces} \textbf{2020}, \emph{12}, 12867--12873\relax
\mciteBstWouldAddEndPuncttrue
\mciteSetBstMidEndSepPunct{\mcitedefaultmidpunct}
{\mcitedefaultendpunct}{\mcitedefaultseppunct}\relax
\EndOfBibitem
\bibitem[Leoncini \latin{et~al.}(2021)Leoncini, Giannuzzi, Giuri, Colella, Listorti, Maiorano, Rizzo, Gigli, and Gambino]{leoncini_electronic_2021}
Leoncini,~M.; Giannuzzi,~R.; Giuri,~A.; Colella,~S.; Listorti,~A.; Maiorano,~V.; Rizzo,~A.; Gigli,~G.; Gambino,~S. Electronic transport, ionic activation energy and trapping phenomena in a polymer-hybrid halide perovskite composite. \emph{Journal of Science: Advanced Materials and Devices} \textbf{2021}, \emph{6}, 543--550\relax
\mciteBstWouldAddEndPuncttrue
\mciteSetBstMidEndSepPunct{\mcitedefaultmidpunct}
{\mcitedefaultendpunct}{\mcitedefaultseppunct}\relax
\EndOfBibitem
\bibitem[Moghadamzadeh \latin{et~al.}(2020)Moghadamzadeh, Hossain, Jakoby, Abdollahi~Nejand, Rueda-Delgado, Schwenzer, Gharibzadeh, Abzieher, Khan, Haghighirad, Howard, Richards, Lemmer, and Paetzold]{moghadamzadeh_spontaneous_2020}
Moghadamzadeh,~S.; Hossain,~I.~M.; Jakoby,~M.; Abdollahi~Nejand,~B.; Rueda-Delgado,~D.; Schwenzer,~J.~A.; Gharibzadeh,~S.; Abzieher,~T.; Khan,~M.~R.; Haghighirad,~A.~A.; Howard,~I.~A.; Richards,~B.~S.; Lemmer,~U.; Paetzold,~U.~W. Spontaneous enhancement of the stable power conversion efficiency in perovskite solar cells. \emph{Journal of Materials Chemistry A} \textbf{2020}, \emph{8}, 670--682\relax
\mciteBstWouldAddEndPuncttrue
\mciteSetBstMidEndSepPunct{\mcitedefaultmidpunct}
{\mcitedefaultendpunct}{\mcitedefaultseppunct}\relax
\EndOfBibitem
\bibitem[Ciavatti \latin{et~al.}(2024)Ciavatti, Foderà, Armaroli, Maserati, Colantoni, Fraboni, and Cavalcoli]{ciavatti_radiation_2024}
Ciavatti,~A.; Foderà,~V.; Armaroli,~G.; Maserati,~L.; Colantoni,~E.; Fraboni,~B.; Cavalcoli,~D. Radiation {Hardness} and {Defects} {Activity} in {PEA}$_{\textrm{2}}$ {PbBr}$_{\textrm{4}}$ {Single} {Crystals}. \emph{Advanced Functional Materials} \textbf{2024}, \emph{34}, 2405291\relax
\mciteBstWouldAddEndPuncttrue
\mciteSetBstMidEndSepPunct{\mcitedefaultmidpunct}
{\mcitedefaultendpunct}{\mcitedefaultseppunct}\relax
\EndOfBibitem
\bibitem[Seid \latin{et~al.}(2024)Seid, Sarisozen, Peña‐Camargo, Ozen, Gutierrez‐Partida, Solano, Steele, Stolterfoht, Neher, and Lang]{seid_understanding_2024}
Seid,~B.~A.; Sarisozen,~S.; Peña‐Camargo,~F.; Ozen,~S.; Gutierrez‐Partida,~E.; Solano,~E.; Steele,~J.~A.; Stolterfoht,~M.; Neher,~D.; Lang,~F. Understanding and {Mitigating} {Atomic} {Oxygen}‐{Induced} {Degradation} of {Perovskite} {Solar} {Cells} for {Near}‐{Earth} {Space} {Applications}. \emph{Small} \textbf{2024}, \emph{20}, 2311097\relax
\mciteBstWouldAddEndPuncttrue
\mciteSetBstMidEndSepPunct{\mcitedefaultmidpunct}
{\mcitedefaultendpunct}{\mcitedefaultseppunct}\relax
\EndOfBibitem
\bibitem[Pallotta \latin{et~al.}(2025)Pallotta, Faini, Toniolo, Larini, Schmidt, Marras, Pica, Cavalli, Mattioni, Hueso, Degani, Martin-Garcia, Ehrler, and Grancini]{pallotta_reducing_2025}
Pallotta,~R.; Faini,~F.; Toniolo,~F.; Larini,~V.; Schmidt,~M.; Marras,~S.; Pica,~G.; Cavalli,~S.; Mattioni,~S.; Hueso,~L.; Degani,~M.; Martin-Garcia,~B.; Ehrler,~B.; Grancini,~G. Reducing the MAPbI3 microstrain by fast crystallization. \emph{Joule} \textbf{2025}, \emph{9}, 101964: 1--10\relax
\mciteBstWouldAddEndPuncttrue
\mciteSetBstMidEndSepPunct{\mcitedefaultmidpunct}
{\mcitedefaultendpunct}{\mcitedefaultseppunct}\relax
\EndOfBibitem
\bibitem[Kim \latin{et~al.}(2020)Kim, Kim, and Park]{kim_first-principles_2020}
Kim,~B.; Kim,~J.; Park,~N. First-principles identification of the charge-shifting mechanism and ferroelectricity in hybrid halide perovskites. \emph{Scientific Reports} \textbf{2020}, \emph{10}, 19635\relax
\mciteBstWouldAddEndPuncttrue
\mciteSetBstMidEndSepPunct{\mcitedefaultmidpunct}
{\mcitedefaultendpunct}{\mcitedefaultseppunct}\relax
\EndOfBibitem
\bibitem[Whitfield \latin{et~al.}(2016)Whitfield, Herron, Guise, Page, Cheng, Milas, and Crawford]{whitfield_structures_2016}
Whitfield,~P.~S.; Herron,~N.; Guise,~W.~E.; Page,~K.; Cheng,~Y.~Q.; Milas,~I.; Crawford,~M.~K. Structures, {Phase} {Transitions} and {Tricritical} {Behavior} of the {Hybrid} {Perovskite} {Methyl} {Ammonium} {Lead} {Iodide}. \emph{Scientific Reports} \textbf{2016}, \emph{6}, 35685\relax
\mciteBstWouldAddEndPuncttrue
\mciteSetBstMidEndSepPunct{\mcitedefaultmidpunct}
{\mcitedefaultendpunct}{\mcitedefaultseppunct}\relax
\EndOfBibitem
\bibitem[Futscher and Milić(2021)Futscher, and Milić]{futscher_mixed_2021}
Futscher,~M.~H.; Milić,~J.~V. Mixed {Conductivity} of {Hybrid} {Halide} {Perovskites}: {Emerging} {Opportunities} and {Challenges}. \emph{Frontiers in Energy Research} \textbf{2021}, \emph{9}, 629074\relax
\mciteBstWouldAddEndPuncttrue
\mciteSetBstMidEndSepPunct{\mcitedefaultmidpunct}
{\mcitedefaultendpunct}{\mcitedefaultseppunct}\relax
\EndOfBibitem
\bibitem[Shah \latin{et~al.}(2024)Shah, Yang, Köhnen, Ugur, Khenkin, Thiesbrummel, Li, Holte, Berwig, Scherler, Forozi, Diekmann, Peña‐Camargo, Remec, Kalasariya, Aydin, Lang, Snaith, Neher, De~Wolf, Ulbrich, Albrecht, and Stolterfoht]{shah_impact_2024}
Shah,~S. \latin{et~al.}  Impact of {Ion} {Migration} on the {Performance} and {Stability} of {Perovskite}‐{Based} {Tandem} {Solar} {Cells}. \emph{Advanced Energy Materials} \textbf{2024}, \emph{14}, 2400720\relax
\mciteBstWouldAddEndPuncttrue
\mciteSetBstMidEndSepPunct{\mcitedefaultmidpunct}
{\mcitedefaultendpunct}{\mcitedefaultseppunct}\relax
\EndOfBibitem
\bibitem[Jacobs \latin{et~al.}(2018)Jacobs, Shen, Pfeffer, Peng, White, Beck, and Catchpole]{jacobs_two_2018}
Jacobs,~D.~A.; Shen,~H.; Pfeffer,~F.; Peng,~J.; White,~T.~P.; Beck,~F.~J.; Catchpole,~K.~R. The two faces of capacitance: {New} interpretations for electrical impedance measurements of perovskite solar cells and their relation to hysteresis. \emph{Journal of Applied Physics} \textbf{2018}, \emph{124}, 225702\relax
\mciteBstWouldAddEndPuncttrue
\mciteSetBstMidEndSepPunct{\mcitedefaultmidpunct}
{\mcitedefaultendpunct}{\mcitedefaultseppunct}\relax
\EndOfBibitem
\bibitem[Birkhold \latin{et~al.}(2018)Birkhold, Precht, Giridharagopal, Eperon, Schmidt-Mende, and Ginger]{birkhold_direct_2018}
Birkhold,~S.~T.; Precht,~J.~T.; Giridharagopal,~R.; Eperon,~G.~E.; Schmidt-Mende,~L.; Ginger,~D.~S. Direct {Observation} and {Quantitative} {Analysis} of {Mobile} {Frenkel} {Defects} in {Metal} {Halide} {Perovskites} {Using} {Scanning} {Kelvin} {Probe} {Microscopy}. \emph{The Journal of Physical Chemistry C} \textbf{2018}, \emph{122}, 12633--12639\relax
\mciteBstWouldAddEndPuncttrue
\mciteSetBstMidEndSepPunct{\mcitedefaultmidpunct}
{\mcitedefaultendpunct}{\mcitedefaultseppunct}\relax
\EndOfBibitem
\bibitem[Jacobs \latin{et~al.}(2022)Jacobs, Wolff, Chin, Artuk, Ballif, and Jeangros]{jacobs_lateral_2022}
Jacobs,~D.~A.; Wolff,~C.~M.; Chin,~X.-Y.; Artuk,~K.; Ballif,~C.; Jeangros,~Q. Lateral ion migration accelerates degradation in halide perovskite devices. \emph{Energy \& Environmental Science} \textbf{2022}, \emph{15}, 5324--5339\relax
\mciteBstWouldAddEndPuncttrue
\mciteSetBstMidEndSepPunct{\mcitedefaultmidpunct}
{\mcitedefaultendpunct}{\mcitedefaultseppunct}\relax
\EndOfBibitem
\bibitem[Li \latin{et~al.}(2018)Li, Guerrero, Huettner, and Bisquert]{li_unravelling_2018}
Li,~C.; Guerrero,~A.; Huettner,~S.; Bisquert,~J. Unravelling the role of vacancies in lead halide perovskite through electrical switching of photoluminescence. \emph{Nature Communications} \textbf{2018}, \emph{9}, 5113\relax
\mciteBstWouldAddEndPuncttrue
\mciteSetBstMidEndSepPunct{\mcitedefaultmidpunct}
{\mcitedefaultendpunct}{\mcitedefaultseppunct}\relax
\EndOfBibitem
\bibitem[McCallum \latin{et~al.}(2024)McCallum, Nicholls, Jensen, Cowley, Lerpinière, and Walker]{mccallum_bayesian_2024}
McCallum,~S.~G.; Nicholls,~O.; Jensen,~K.~O.; Cowley,~M.~V.; Lerpinière,~J.~E.; Walker,~A.~B. Bayesian parameter estimation for characterising mobile ion vacancies in perovskite solar cells. \emph{Journal of Physics: Energy} \textbf{2024}, \emph{6}, 015005\relax
\mciteBstWouldAddEndPuncttrue
\mciteSetBstMidEndSepPunct{\mcitedefaultmidpunct}
{\mcitedefaultendpunct}{\mcitedefaultseppunct}\relax
\EndOfBibitem
\bibitem[Schmidt \latin{et~al.}(2024)Schmidt, Alvarez, De~Boer, Van De~Ven, and Ehrler]{schmidt_consistent_2024}
Schmidt,~M.~C.; Alvarez,~A.~O.; De~Boer,~J.~J.; Van De~Ven,~L.~J.; Ehrler,~B. Consistent {Interpretation} of {Time}- and {Frequency}-{Domain} {Traces} of {Ion} {Migration} in {Perovskite} {Semiconductors}. \emph{ACS Energy Letters} \textbf{2024}, 5850--5858\relax
\mciteBstWouldAddEndPuncttrue
\mciteSetBstMidEndSepPunct{\mcitedefaultmidpunct}
{\mcitedefaultendpunct}{\mcitedefaultseppunct}\relax
\EndOfBibitem
\bibitem[Futscher \latin{et~al.}(2019)Futscher, Lee, McGovern, Muscarella, Wang, Haider, Fakharuddin, Schmidt-Mende, and Ehrler]{futscher_quantification_2019}
Futscher,~M.~H.; Lee,~J.~M.; McGovern,~L.; Muscarella,~L.~A.; Wang,~T.; Haider,~M.~I.; Fakharuddin,~A.; Schmidt-Mende,~L.; Ehrler,~B. Quantification of ion migration in {CH}$_{\textrm{3}}${NH}$_{\textrm{3}}${PbI}$_{\textrm{3}}$ perovskite solar cells by transient capacitance measurements. \emph{Materials Horizons} \textbf{2019}, \emph{6}, 1497--1503\relax
\mciteBstWouldAddEndPuncttrue
\mciteSetBstMidEndSepPunct{\mcitedefaultmidpunct}
{\mcitedefaultendpunct}{\mcitedefaultseppunct}\relax
\EndOfBibitem
\bibitem[Eames \latin{et~al.}(2015)Eames, Frost, Barnes, O’Regan, Walsh, and Islam]{eames_ionic_2015}
Eames,~C.; Frost,~J.~M.; Barnes,~P. R.~F.; O’Regan,~B.~C.; Walsh,~A.; Islam,~M.~S. Ionic transport in hybrid lead iodide perovskite solar cells. \emph{Nature Communications} \textbf{2015}, \emph{6}, 7497\relax
\mciteBstWouldAddEndPuncttrue
\mciteSetBstMidEndSepPunct{\mcitedefaultmidpunct}
{\mcitedefaultendpunct}{\mcitedefaultseppunct}\relax
\EndOfBibitem
\bibitem[Haruyama \latin{et~al.}(2015)Haruyama, Sodeyama, Han, and Tateyama]{haruyama_first-principles_2015}
Haruyama,~J.; Sodeyama,~K.; Han,~L.; Tateyama,~Y. First-{Principles} {Study} of {Ion} {Diffusion} in {Perovskite} {Solar} {Cell} {Sensitizers}. \emph{Journal of the American Chemical Society} \textbf{2015}, \emph{137}, 10048--10051\relax
\mciteBstWouldAddEndPuncttrue
\mciteSetBstMidEndSepPunct{\mcitedefaultmidpunct}
{\mcitedefaultendpunct}{\mcitedefaultseppunct}\relax
\EndOfBibitem
\bibitem[Jong \latin{et~al.}(2018)Jong, Yu, Ri, McMahon, Harrison, Barnes, and Walsh]{jong_influence_2018}
Jong,~U.-G.; Yu,~C.-J.; Ri,~G.-C.; McMahon,~A.~P.; Harrison,~N.; Barnes,~P. R.~F.; Walsh,~A. Influence of water intercalation and hydration on chemical decomposition and ion transport in methylammonium lead halide perovskites. \emph{Journal of Materials Chemistry A} \textbf{2018}, \emph{6}, 1067--1074\relax
\mciteBstWouldAddEndPuncttrue
\mciteSetBstMidEndSepPunct{\mcitedefaultmidpunct}
{\mcitedefaultendpunct}{\mcitedefaultseppunct}\relax
\EndOfBibitem
\bibitem[Shao \latin{et~al.}(2016)Shao, Fang, Li, Wang, Dong, Deng, Yuan, Wei, Wang, Gruverman, Shield, and Huang]{shao_grain_2016}
Shao,~Y.; Fang,~Y.; Li,~T.; Wang,~Q.; Dong,~Q.; Deng,~Y.; Yuan,~Y.; Wei,~H.; Wang,~M.; Gruverman,~A.; Shield,~J.; Huang,~J. Grain boundary dominated ion migration in polycrystalline organic–inorganic halide perovskite films. \emph{Energy \& Environmental Science} \textbf{2016}, \emph{9}, 1752--1759\relax
\mciteBstWouldAddEndPuncttrue
\mciteSetBstMidEndSepPunct{\mcitedefaultmidpunct}
{\mcitedefaultendpunct}{\mcitedefaultseppunct}\relax
\EndOfBibitem
\bibitem[Ghasemi \latin{et~al.}(2023)Ghasemi, Guo, Darabi, Wang, Wang, Huang, Lefler, Taussig, Chauhan, Baucom, Kim, Gomez, Atkin, Priya, and Amassian]{ghasemi_multiscale_2023}
Ghasemi,~M.; Guo,~B.; Darabi,~K.; Wang,~T.; Wang,~K.; Huang,~C.-W.; Lefler,~B.~M.; Taussig,~L.; Chauhan,~M.; Baucom,~G.; Kim,~T.; Gomez,~E.~D.; Atkin,~J.~M.; Priya,~S.; Amassian,~A. A multiscale ion diffusion framework sheds light on the diffusion–stability–hysteresis nexus in metal halide perovskites. \emph{Nature Materials} \textbf{2023}, \emph{22}, 329--337\relax
\mciteBstWouldAddEndPuncttrue
\mciteSetBstMidEndSepPunct{\mcitedefaultmidpunct}
{\mcitedefaultendpunct}{\mcitedefaultseppunct}\relax
\EndOfBibitem
\bibitem[Wei \latin{et~al.}(2018)Wei, Ma, Wang, Dou, Cui, Huang, Ji, Jia, Jia, Sajid, Elseman, Chu, Li, Jiang, Qiao, Yuan, and Li]{wei_ion-migration_2018}
Wei,~D. \latin{et~al.}  Ion-{Migration} {Inhibition} by the {Cation}-π {Interaction} in {Perovskite} {Materials} for {Efficient} and {Stable} {Perovskite} {Solar} {Cells}. \emph{Advanced Materials} \textbf{2018}, \emph{30}, 1707583\relax
\mciteBstWouldAddEndPuncttrue
\mciteSetBstMidEndSepPunct{\mcitedefaultmidpunct}
{\mcitedefaultendpunct}{\mcitedefaultseppunct}\relax
\EndOfBibitem
\bibitem[Weber \latin{et~al.}(2018)Weber, Hermes, Turren-Cruz, Gort, Bergmann, Gilson, Hagfeldt, Graetzel, Tress, and Berger]{weber_how_2018}
Weber,~S. A.~L.; Hermes,~I.~M.; Turren-Cruz,~S.-H.; Gort,~C.; Bergmann,~V.~W.; Gilson,~L.; Hagfeldt,~A.; Graetzel,~M.; Tress,~W.; Berger,~R. How the formation of interfacial charge causes hysteresis in perovskite solar cells. \emph{Energy \& Environmental Science} \textbf{2018}, \emph{11}, 2404--2413\relax
\mciteBstWouldAddEndPuncttrue
\mciteSetBstMidEndSepPunct{\mcitedefaultmidpunct}
{\mcitedefaultendpunct}{\mcitedefaultseppunct}\relax
\EndOfBibitem
\bibitem[Yang \latin{et~al.}(2016)Yang, Ming, Shi, Zhang, and Du]{yang_fast_2016}
Yang,~D.; Ming,~W.; Shi,~H.; Zhang,~L.; Du,~M.-H. Fast {Diffusion} of {Native} {Defects} and {Impurities} in {Perovskite} {Solar} {Cell} {Material} {CH}$_{\textrm{3}}$ {NH}$_{\textrm{3}}$ {PbI}$_{\textrm{3}}$. \emph{Chemistry of Materials} \textbf{2016}, \emph{28}, 4349--4357\relax
\mciteBstWouldAddEndPuncttrue
\mciteSetBstMidEndSepPunct{\mcitedefaultmidpunct}
{\mcitedefaultendpunct}{\mcitedefaultseppunct}\relax
\EndOfBibitem
\bibitem[Pitaro \latin{et~al.}(2022)Pitaro, Tekelenburg, Shao, and Loi]{pitaro_tin_2022}
Pitaro,~M.; Tekelenburg,~E.~K.; Shao,~S.; Loi,~M.~A. Tin Halide Perovskites: From Fundamental Properties to Solar Cells. \emph{Advanced Materials} \textbf{2022}, \emph{34}, 2105844\relax
\mciteBstWouldAddEndPuncttrue
\mciteSetBstMidEndSepPunct{\mcitedefaultmidpunct}
{\mcitedefaultendpunct}{\mcitedefaultseppunct}\relax
\EndOfBibitem
\bibitem[Bertoluzzi \latin{et~al.}(2020)Bertoluzzi, Boyd, Rolston, Xu, Prasanna, O’Regan, and McGehee]{bertoluzzi_mobile_2020}
Bertoluzzi,~L.; Boyd,~C.~C.; Rolston,~N.; Xu,~J.; Prasanna,~R.; O’Regan,~B.~C.; McGehee,~M.~D. Mobile {Ion} {Concentration} {Measurement} and {Open}-{Access} {Band} {Diagram} {Simulation} {Platform} for {Halide} {Perovskite} {Solar} {Cells}. \emph{Joule} \textbf{2020}, \emph{4}, 109--127\relax
\mciteBstWouldAddEndPuncttrue
\mciteSetBstMidEndSepPunct{\mcitedefaultmidpunct}
{\mcitedefaultendpunct}{\mcitedefaultseppunct}\relax
\EndOfBibitem
\bibitem[Alvarez \latin{et~al.}(2024)Alvarez, García‐Batlle, Lédée, Gros‐Daillon, Guillén, Verilhac, Lemercier, Zaccaro, Marsal, Almora, and Garcia‐Belmonte]{alvarez_ion_2024}
Alvarez,~A.~O.; García‐Batlle,~M.; Lédée,~F.; Gros‐Daillon,~E.; Guillén,~J.~M.; Verilhac,~J.; Lemercier,~T.; Zaccaro,~J.; Marsal,~L.~F.; Almora,~O.; Garcia‐Belmonte,~G. Ion {Migration} and {Space}‐{Charge} {Zones} in {Metal} {Halide} {Perovskites} {Through} {Short}‐{Circuit} {Transient} {Current} and {Numerical} {Simulations}. \emph{Advanced Electronic Materials} \textbf{2024}, 2400241\relax
\mciteBstWouldAddEndPuncttrue
\mciteSetBstMidEndSepPunct{\mcitedefaultmidpunct}
{\mcitedefaultendpunct}{\mcitedefaultseppunct}\relax
\EndOfBibitem
\bibitem[McGovern \latin{et~al.}(2021)McGovern, Grimaldi, Futscher, Hutter, Muscarella, Schmidt, and Ehrler]{mcgovern_reduced_2021}
McGovern,~L.; Grimaldi,~G.; Futscher,~M.~H.; Hutter,~E.~M.; Muscarella,~L.~A.; Schmidt,~M.~C.; Ehrler,~B. Reduced {Barrier} for {Ion} {Migration} in {Mixed}-{Halide} {Perovskites}. \emph{ACS Applied Energy Materials} \textbf{2021}, \emph{4}, 13431--13437\relax
\mciteBstWouldAddEndPuncttrue
\mciteSetBstMidEndSepPunct{\mcitedefaultmidpunct}
{\mcitedefaultendpunct}{\mcitedefaultseppunct}\relax
\EndOfBibitem
\end{mcitethebibliography}
\newpage

\end{document}